\documentclass[zpreprint,zbstepj,british]{./zeus_Kpaper}
\usepackage{./zeusdet_units}
\usepackage{subcaption}
\usepackage{caption}
\usepackage{rotating}
\usepackage{./lineno}
\usepackage{amsmath}
\usepackage{verbatim}
\usepackage{stackengine}
\usepackage{graphicx}
\usepackage{pdfpages}

\chardef\usc=95
\chardef\til=126
\catcode`\@=11 
\DeclareRobustCommand\xdotspace{\futurelet\@let@token\@xdotspace}
\def\@xdotspace{%
  \ifx\@let@token.\else
  \ifx\@let@token\bgroup.\else
  \ifx\@let@token\egroup.\else
  \ifx\@let@token\/.\else
  \ifx\@let@token\ .\else
  \ifx\@let@token~.\else
  \ifx\@let@token!.\else
  \ifx\@let@token,.\else
  \ifx\@let@token:.\else
  \ifx\@let@token;.\else
  \ifx\@let@token?.\else
  \ifx\@let@token/.\else
  \ifx\@let@token'.\else
  \ifx\@let@token).\else
  \ifx\@let@token-.\else
  \ifx\@let@token\@xobeysp.\else
  \ifx\@let@token\space.\else
  \ifx\@let@token\@sptoken.\else
   .\space
   \fi\fi\fi\fi\fi\fi\fi\fi\fi\fi\fi\fi\fi\fi\fi\fi\fi\fi}
\catcode`\@=12 

\newcommand{\stru}[2]{%
   \relax\ifmmode\hbox{\vrule height#1 depth#2 width0pt}%
   \else\vrule height#1 depth#2 width0pt\fi}

\newcommand{\Ronum}[1]{\uppercase\expandafter{\romannumeral#1}}
\newcommand{\ronum}[1]{\expandafter{\romannumeral#1}}
\DeclareRobustCommand{\LaTeXZ}{%
  \LaTeX\kern-.05em4\kern-.1em
  {\raisebox{-0.2ex}{$\scriptstyle\text{ZEUS}$}}\xspace}

\newcommand{\slashfrac}[2]{%
  \raisebox{0.5ex}{\ensuremath #1}\kern-0.12em/\kern-0.08em
  \raisebox{-.8ex}{\ensuremath #2}}

\newcommand{\sqr}[3]{%
    {\vcenter{\hrule height.#3ex\hbox{\vrule width.#2ex height#1ex
     \kern#1ex\vrule width.#3ex}\hrule height.#2ex}}}

\catcode`\@=11 
\newcommand{\parenbar}{\mathpalette\p@renb@r}
\def\p@renb@r#1#2{\vbox{%
  \ifx#1\scriptscriptstyle \dimen@.7em\dimen@ii.2em\else
  \ifx#1\scriptstyle \dimen@.8em\dimen@ii.25em\else
  \dimen@1em\dimen@ii.4em\fi\fi \offinterlineskip
  \ialign{\hfill##\hfill\cr
    \vbox{\hrule width\dimen@ii}\cr
    \noalign{\vskip-.3ex}%
    \hbox to\dimen@{$\mathchar300\hfil\mathchar301$}\cr
    \noalign{\vskip-.3ex}%
    $#1#2$\cr}}}
\catcode`\@=12 

\ifzeusunit
  
\else
  
\fi

\newcommand{\IP}{{\rm I$\kern-0.01667em$P}\xspace}

\mathchardef\qsm=63
\mathchardef\pls=43
\mathchardef\mns=512
\mathchardef\plm=518
\mathchardef\eql=61
\mathchardef\smallleft=300
\mathchardef\smallright=301
\mathchardef\les=316
\mathchardef\gre=318
\mathchardef\leq=532
\mathchardef\grq=533

\catcode`\@=11 
\newcounter{pict@width}
\newcounter{pict@height}
\newlength{\pict@scale}
\newcommand{\psfigadd}[4]{%
\setcounter{pict@width}{1*\ratio{#2+\pict@scale/2}{\pict@scale}}
\setcounter{pict@height}{1*\ratio{#3+\pict@scale/2}{\pict@scale}}
\setlength{\unitlength}{\pict@scale}
\hbox to #2{\hspace{-\fill}\begin{picture}(\thepict@width,\thepict@height)
\put(0,0){\psfig{figure=#1,width=#2,height=#3,clip=}}
\SetScale{0.283466457}
\SetWidth{1.763889}
{#4}
\end{picture}}
}
\newcounter{pict@widthfst}
\newcounter{pict@widthscd}
\newcounter{pict@widthtot}
\newcommand{\psfigaddtwo}[7]{%
\setcounter{pict@widthfst}{1*\ratio{#2+\pict@scale/2}{\pict@scale}}
\setcounter{pict@widthscd}{1*\ratio{#2+#4+\pict@scale/2}{\pict@scale}}
\setcounter{pict@widthtot}{1*\ratio{#2+#4+#6+\pict@scale/2}{\pict@scale}}
\setcounter{pict@height}{1*\ratio{#3+\pict@scale/2}{\pict@scale}}
\setlength{\unitlength}{\pict@scale}
\hbox{\hspace{-\fill}\begin{picture}(\thepict@widthtot,\thepict@height)
\put(0,0){\psfig{figure=#1,width=#2,height=#3,clip=}}
\put(\thepict@widthscd,0){\psfig{figure=#5,width=#6,height=#3,clip=}}
\SetScale{0.283466457}
\SetWidth{1.763889}
{#7}
\end{picture}}
}
\newcommand{\psfigror}[4]{%
\setcounter{pict@width}{1*\ratio{#2+\pict@scale/2}{\pict@scale}}
\setcounter{pict@height}{1*\ratio{#3+\pict@scale/2}{\pict@scale}}
\setlength{\unitlength}{\pict@scale}
\hbox{\begin{picture}(\thepict@width,\thepict@height)
\put(0,\thepict@height){\psfig{figure=#1,width=#3,height=#2,clip=,angle=270}}
\SetScale{0.283466457}
\SetWidth{1.763889}
{#4}
\end{picture}}
}
\newcommand{\psfigrol}[4]{%
\setcounter{pict@width}{1*\ratio{#2+\pict@scale/2}{\pict@scale}}
\setcounter{pict@height}{1*\ratio{#3+\pict@scale/2}{\pict@scale}}
\setlength{\unitlength}{\pict@scale}
\hbox{\begin{picture}(\thepict@width,\thepict@height)
\put(0,0){\psfig{figure=#1,width=#3,height=#2,clip=,angle=90}}
\SetScale{0.283466457}
\SetWidth{1.763889}
{#4}
\end{picture}}
}
\catcode`\@=12 
\newlength\listtextwidth

\catcode`\@=11 
\newlength{\@tabfninsert}
\newlength{\@tabfnwidth}
\newcommand{\tabfootnote}[2]{%
  \setlength{\@tabfninsert}{0.8em}
  \setlength{\@tabfnwidth}{\textwidth}
  \addtolength{\@tabfnwidth}{-\@tabfninsert}
  \addtolength{\@tabfnwidth}{-0.4em}
  \noindent\makebox[\@tabfninsert][r]{\footnotesize$^{#1}$\hfil}\hfill%
  \parbox[t]{\@tabfnwidth}{\footnotesize #2\hfill}}
\catcode`\@=12 

\def\citeCTD{{\cite{%
nim:a279:290,*npps:b32:181,*nim:a338:254%
}}\xspace}
\def\citeMVD{{\cite{%
nim:a581:656%
}}\xspace}
\def\citeSTT{{\cite{%
nim:a535:191%
}}\xspace}
\def\citeCAL{{\cite{%
nim:a309:77,*nim:a309:101,*nim:a321:356,*nim:a336:23%
}}\xspace}
\def\citeHES{{\cite{%
nim:a277:176%
}}\xspace}
\def\citeSixmT{{\cite{%
thesis:gosau:2007%
}}\xspace}
\def\citeBAC{{\cite{%
nim:a300:480%
}}\xspace}
\def\citePCAL{{\cite{%
desy-92-066,*zfp:c63:391,*acpp:b32:2025%
}}\xspace}
\def\citeSPECTRO{{\cite{%
nim:a565:572%
}}\xspace}

\begin{document}
\prepnum{DESY--19--054}
\prepdate{March 2019}

\zeustitle{
  Charm production in charged current deep inelastic scattering at HERA
}

\zeusauthor{ZEUS Collaboration}

\maketitle

\begin{abstract}\noindent
Charm production in charged current deep inelastic scattering has been measured
for the first time in $e^{\pm}p$ collisions, using data collected with the ZEUS
detector at HERA, corresponding to an integrated luminosity of $358 \pb^{-1}$.
Results are presented separately for $e^{+}p$ and $e^{-}p$ scattering at a
centre-of-mass energy of $\sqrt{s} = 318 \gev$ within a kinematic phase-space
region of $200\gev^{2}<Q^{2}<60000\gev^{2}$ and $y<0.9$, where $Q^{2}$ is the
squared four-momentum transfer and $y$ is the inelasticity. The measured cross
sections of electroweak charm production are consistent with expectations from
the Standard Model within the large statistical uncertainties.
\end{abstract}

\thispagestyle{empty}
\cleardoublepage


# The ZEUS Collaboration

I. Abt[19], L. Adamczyk[7], R. Aggarwal[3,b], V. Aushev[17], O. Behnke[9], U. Behrens[9], A. Bertolin[21],
I. Bloch[10], I. Brock[2], N.H. Brook[28,m], R. Brugnera[22], A. Bruni[1], P.J. Bussey[11], A. Caldwell[19],
M. Capua[4], C.D. Catterall[32], J. Chwastowski[6], J. Ciborowski[29,o], R. Ciesielski[9,d],
A.M. Cooper-Sarkar[20], M. Corradi[1,a], R.K. Dementiev[18], S. Dusini[21], J. Ferrando[9],
B. Foster[13,j], E. Gallo[13,k], D. Gangadharan[14], A. Garfagnini[22], A. Geiser[9], L.K. Gladilin[18],
Yu.A. Golubkov[18], G. Grzelak[29], C. Gwenlan[20], D. Hochman[31], N.Z. Jomhari[9], I. Kadenko[17],
S. Kananov[23], U. Karshon[31], P. Kaur[3,c], R. Klanner[13], U. Klein[9,e], I.A. Korzhavina[18],
N. Kovalchuk[13], H. Kowalski[9], O. Kuprash[9,f], M. Kuze[25], B.B. Levchenko[18], A. Levy[23],
B. Löhr[9], E. Lohrmann[13], A. Longhin[22], O.Yu. Lukina[18], I. Makarenko[9], J. Malka[9,g],
S. Masciocchi[12,i], K. Nagano[15], J.D. Nam[24], J. Onderwaater[14,l], Yu. Onishchuk[17], E. Paul[2],
I. Pidhurskyi[17], A. Polini[1], M. Przybycień[7], A. Quintero[24], M. Ruspa[27], D.H. Saxon[11],
U. Schneekloth[9], T. Schörner-Sadenius[9], I. Selyuzhenkov[12], M. Shchedrolosiev[17],
L.M. Shcheglova[18], O. Shkola[17], I.O. Skillicorn[11], W. Słomiński[8], A. Solano[26], L. Stanco[21],
N. Stefaniuk[9], P. Stopa[6], B. Surrow[24], J. Sztuk-Dambietz[13,g], E. Tassi[4], K. Tokushuku[15],
M. Turcato[13,g], O. Turkot[9], T. Tymieniecka[30], A. Verbytskyi[19], W.A.T. Wan Abdullah[5],
K. Wichmann[9], M. Wing[28,n], S. Yamada[15], Y. Yamazaki[16], A.F. Żarnecki[29], L. Zawiejski[6],
O. Zenaiev[9,h]

1.  INFN Bologna, Bologna, Italy [A]
2.  Physikalisches Institut der Universität Bonn, Bonn, Germany [B]
3.  Panjab University, Department of Physics, Chandigarh, India
4.  Calabria University, Physics Department and INFN, Cosenza, Italy [A]
5.  National Centre for Particle Physics, Universiti Malaya, 50603 Kuala Lumpur, Malaysia [C]
6.  The Henryk Niewodniczanski Institute of Nuclear Physics, Polish Academy of Sciences, Krakow, Poland
7.  AGH University of Science and Technology, Faculty of Physics and Applied Computer Science, Krakow, Poland
8.  Department of Physics, Jagellonian University, Krakow, Poland [D]
9.  Deutsches Elektronen-Synchrotron DESY, Hamburg, Germany
10. Deutsches Elektronen-Synchrotron DESY, Zeuthen, Germany
11. School of Physics and Astronomy, University of Glasgow, Glasgow, United Kingdom [E]
12. GSI Helmholtzzentrum für Schwerionenforschung GmbH, Darmstadt, Germany
13. Hamburg University, Institute of Experimental Physics, Hamburg, Germany [F]
14. Physikalisches Institut of the University of Heidelberg, Heidelberg, Germany
15. Institute of Particle and Nuclear Studies, KEK, Tsukuba, Japan [G]
16. Department of Physics, Kobe University, Kobe, Japan [G]
17. Department of Nuclear Physics, National Taras Shevchenko University of Kyiv, Kyiv, Ukraine
18. Lomonosov Moscow State University, Skobeltsyn Institute of Nuclear Physics, Moscow, Russia
19. Max-Planck-Institut für Physik, München, Germany
20. Department of Physics, University of Oxford, Oxford, United Kingdom [E]
21. INFN Padova, Padova, Italy [A]
22. Dipartimento di Fisica e Astronomia dell' Università and INFN, Padova, Italy [A]
23. Raymond and Beverly Sackler Faculty of Exact Sciences, School of Physics, Tel Aviv University, Tel Aviv, Israel [H]
24. Department of Physics, Temple University, Philadelphia, PA 19122, USA [I]
25. Department of Physics, Tokyo Institute of Technology, Tokyo, Japan [G]
26. Università di Torino and INFN, Torino, Italy [A]
27. Università del Piemonte Orientale, Novara, and INFN, Torino, Italy [A]
28. Physics and Astronomy Department, University College London, London, United Kingdom [E]
29. Faculty of Physics, University of Warsaw, Warsaw, Poland
30. National Centre for Nuclear Research, Warsaw, Poland
31. Department of Particle Physics and Astrophysics, Weizmann Institute, Rehovot, Israel
32. Department of Physics, York University, Ontario, Canada M3J 1P3 [J]



[a] now at INFN Roma, Italy

[b] now at DST-Inspire Faculty, Department of Technology, SPPU, India

[c] now at Sant Longowal Institute of Engineering and Technology, Longowal, Punjab, India

[d] now at Rockefeller University, New York, NY 10065, USA

[e] now at University of Liverpool, United Kingdom

[f] now at University of Freiburg, Freiburg, Germany

[g] now at European X-ray Free-Electron Laser facility GmbH, Hamburg, Germany

[h] now at Hamburg University, Hamburg, Germany

[i] also at Physikalisches Institut of the University of Heidelberg, Heidelberg, Germany

[j] Alexander von Humboldt Professor; also at DESY and University of Oxford

[k] also at DESY

[l] also at GSI Helmholtzzentrum für Schwerionenforschung GmbH, Darmstadt, Germany

[m] now at University of Bath, United Kingdom

[n] also supported by DESY

[o] also at Lodz University, Poland


\pagenumbering{arabic}
\section{Introduction}
\label{sec-int}
\setlength{\footnotesep}{0.5 cm}
Measurements of heavy-flavour production serve as a good testing ground to investigate the predictive power of
perturbative quantum chromodynamics (pQCD) as the large mass provides a natural hard scale.
While charm production in neutral current deep inelastic scattering (NC DIS) and in photoproduction has been extensively
studied at HERA, it has not been measured in charged current deep inelastic scattering (CC DIS) owing to its small cross section. 

In CC DIS, single charm quarks in the final state already occur at the level of the Quark Parton Model (QPM) when either an
incoming $s$ or $d$ quark is converted to a charm quark, or an incoming charm quark is converted to an $s$ or $d$ quark,
as illustrated in Fig.~\ref{fig-subprocess1}  (i, ii). In the latter case, the single charm in the event arises from the
associated charm quark in the proton remnant. In addition, single charm can arise from boson--gluon fusion (BGF) producing a
$c\bar{s}~(c\bar{d})$ quark pair. In this case, the incoming virtual $W$ boson fuses with a gluon from the proton. The gluon
splits into a $s\bar{s}~(d\bar{d})$ or $c\bar{c}$ pair in the initial state, as shown in Fig.~\ref{fig-subprocess1} (iii, iv).
All these $e^{+}p$ processes lead to the same final state, $e^{+}p \rightarrow \bar{\nu}_{e}~c\bar{s}(\bar{d})~X;$
this is also true for $e^{-}p$, $e^{-}p \rightarrow \nu_{e}~\bar{c} s(d)~X.$
The characteristics of the events associated with these subprocesses and their association to particular kinematic
configurations in the final state depend on the QCD scheme chosen, as detailed in the next section.
The subprocess depicted in Fig.~\ref{fig-subprocess1} (i) is directly sensitive to the strange-quark content of the proton
and can be used to constrain it. However, the extraction of the relevant part of the cross section is model dependent.

In the SU(3) flavour model, a perfect symmetry is assumed between the three light flavours, which results in equal quark
densities for the sea quark components in nucleons. This symmetry is broken if the strange-quark density is suppressed by
the mass of the strange quark, as happens in the well established strange-quark suppression in fragmentation~\cite{pdg1}.
This symmetry breaking can also occur in the initial state, depending on $x$, the fraction of the proton momentum
carried by the interacting parton. 
For larger values of $x$, some support for this has been found experimentally, such as in dimuon production in charged
current by the CCFR~\cite{nutev} and NuTeV~\cite{Goncharov:2001qe}, as well as the NOMAD~\cite{Samoylov:2013xoa} and
CHORUS~\cite{KayisTopaksu:2011mx} neutrino scattering experiments. 
However, the interpretation of these measurements depends on nuclear
corrections and charm fragmentation and
no consensus has emerged on the exact level of suppression as a function of $x$.
Additionally, the recent high-precision measurements of inclusive $W$ and $Z$ production by the ATLAS collaboration~\cite{atlas}
report an unsuppressed strange sea in the low-$x$ regime.
A similar result was obtained in a combined global QCD analysis of inclusive $W$ and $Z$ data from both the ATLAS
and CMS experiments~\cite{cskk}. This observation was also supported by the analysis of the
ATLAS $W+c$ data~\cite{atlas2}. However, the CMS $W+c$ data~\cite{Chatrchyan:2013uja,cms} favour
strangeness suppression also at low $x$.
A re-evaluation of the LHC inclusive and $W + c$ measurements and the neutrino scattering measurements by
NOMAD~\cite{Samoylov:2013xoa} and CHORUS~\cite{KayisTopaksu:2011mx} has been
performed~\cite{Alekhin:2014sya,Alekhin:2018}, partly in an attempt to reconcile the factor-of-two discrepancy
in the measured strange-quark densities. The resulting strange-quark parton distribution function (PDF) was reported
to be inconsistent with the ATLAS fit~\cite{atlas}.

This paper presents measurements of charm production in CC DIS in $e^{\pm}p$ collisions using data from
the HERA II data-taking period.
The electroweak contribution to charm-production cross sections is compared with several QCD schemes that are detailed
in the following section.

\section{Charm production in CC DIS at HERA}
\label{sec-charm-cc}

The kinematics of lepton--proton scattering can be described in terms of the Lorentz-invariant variables $x_\textrm{Bj}$, $y$ and $Q^{2}$.
The variable $Q^{2}$ is the negative squared four-momentum of the exchange boson $-q^{2} = -(k-k')^{2}$,
where $k$ and $k'$ are the four-momenta of the incoming and outgoing lepton, respectively.
The Bjorken-$x$ scaling variable, $x_\textrm{Bj}$, is defined as $x_\textrm{Bj} = Q^{2}/(2 p \cdot q)$,
where $p$ is the four-momentum of the incoming proton.
The variable $y$ is the inelasticity defined as $y = Q^{2}/(sx_\textrm{Bj})$,
where $s$ is the squared centre-of-mass energy of the collision.

\begin{figure}[p]
\vfill
\begin{subfigure}{0.5\textwidth}
\phantomcaption
\stackinset{l}{1.7cm}{b}{3.3cm}{(\roman{subfigure})}
{\includegraphics[width=8cm,height=6.7cm]{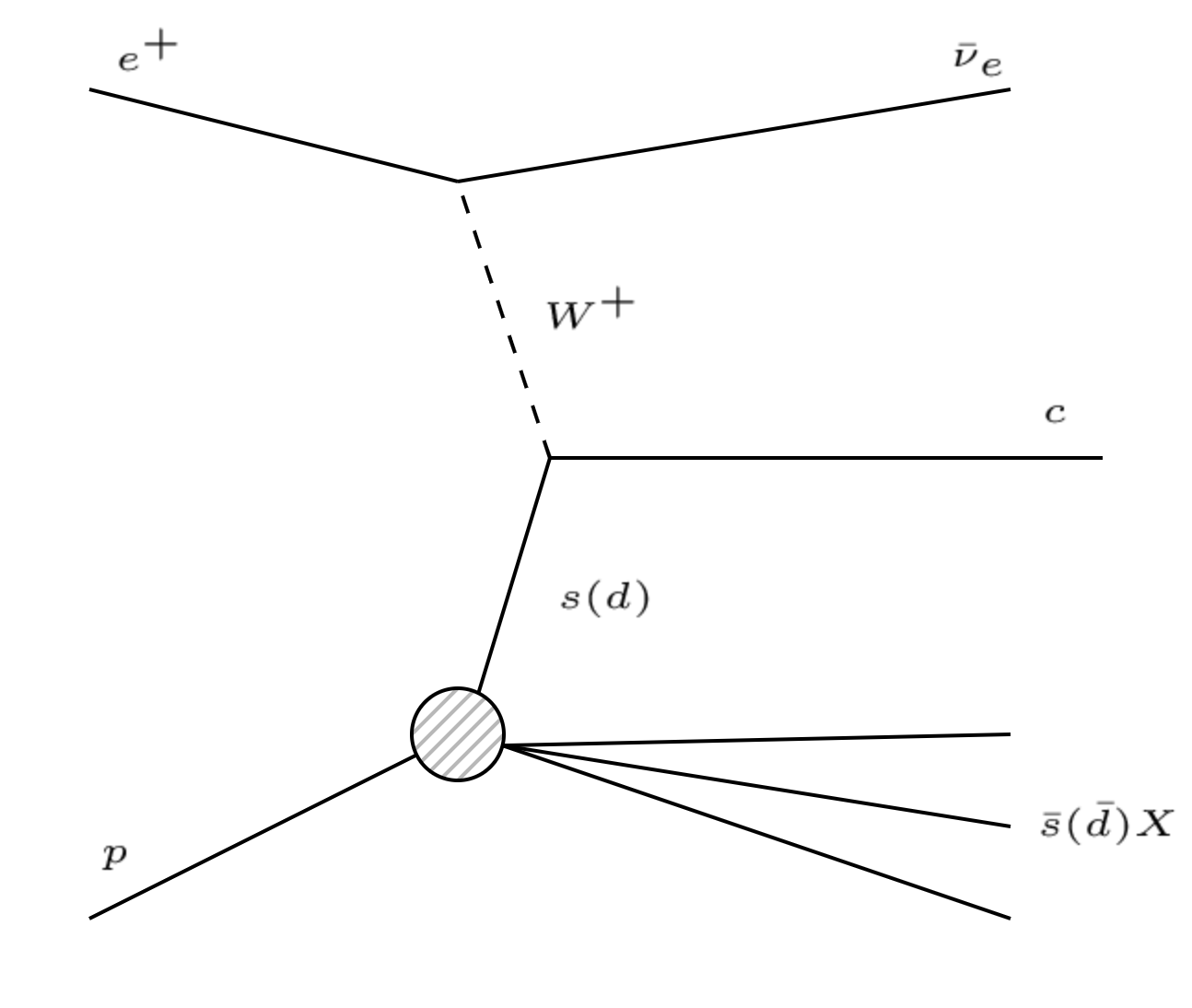}}
\label{fig:subim1}
\end{subfigure}
\begin{subfigure}{0.5\textwidth}
\phantomcaption
\stackinset{l}{1.7cm}{b}{3.3cm}{(\roman{subfigure})}
{\includegraphics[width=8cm,height=6.7cm]{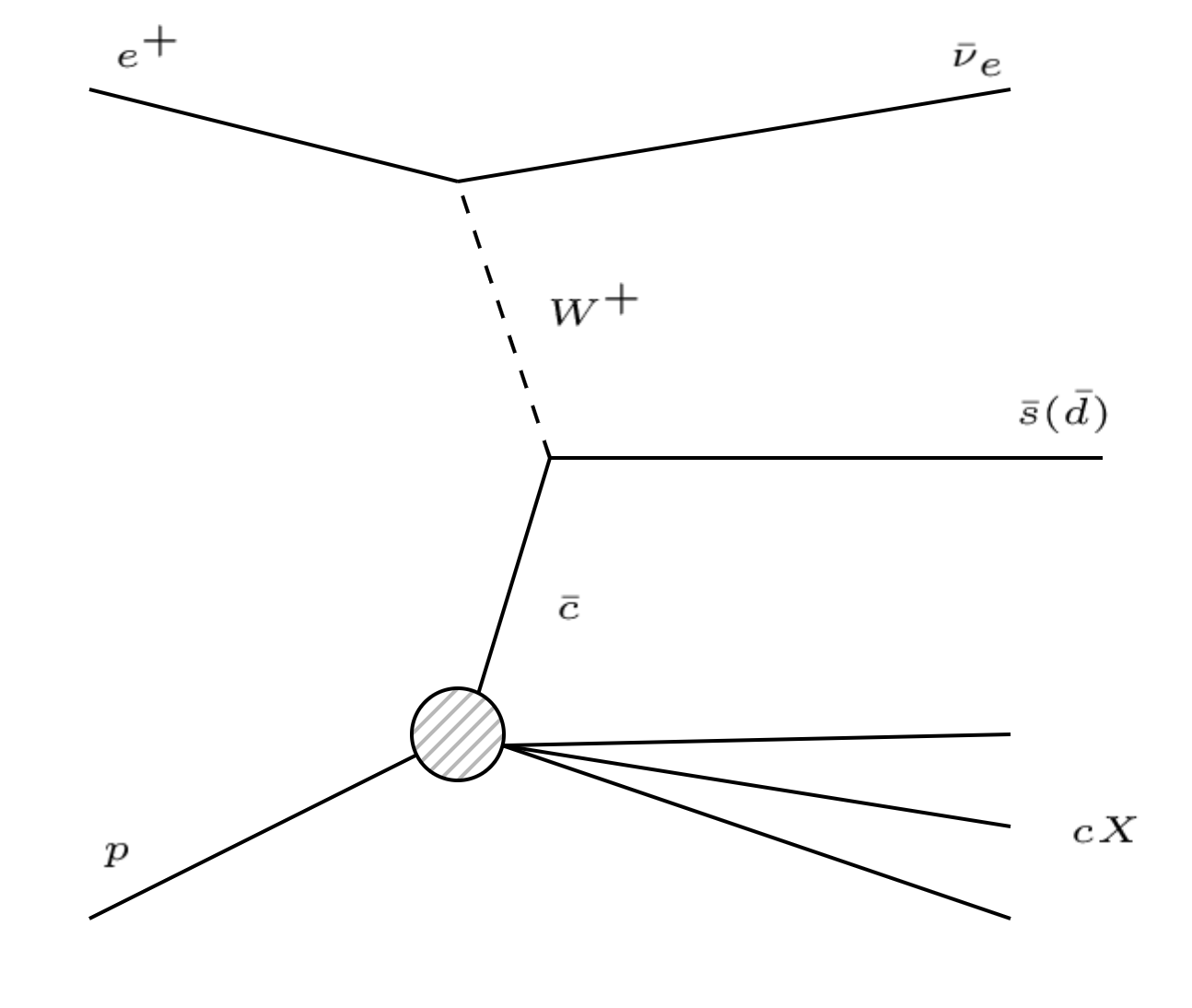}}
\label{fig:subim2}
\end{subfigure}
\begin{subfigure}{0.5\textwidth}
\phantomcaption
\stackinset{l}{1.7cm}{b}{3.3cm}{(\roman{subfigure})}
{\includegraphics[width=8cm,height=6.7cm]{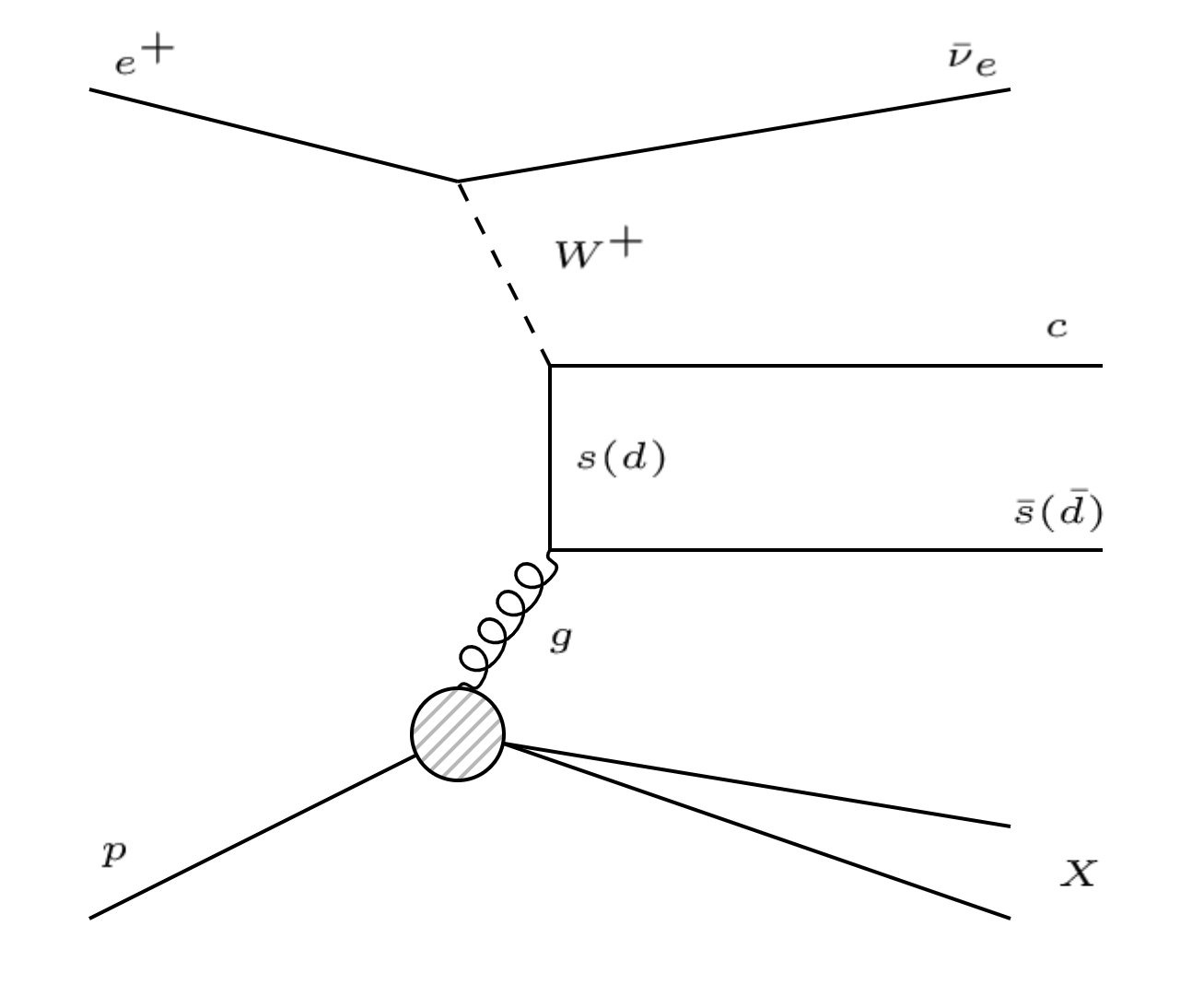}}
\label{fig:subim3}
\end{subfigure}
\begin{subfigure}{0.5\textwidth}
\phantomcaption
\stackinset{l}{1.7cm}{b}{3.3cm}{(\roman{subfigure})}
{\includegraphics[width=8cm,height=6.7cm]{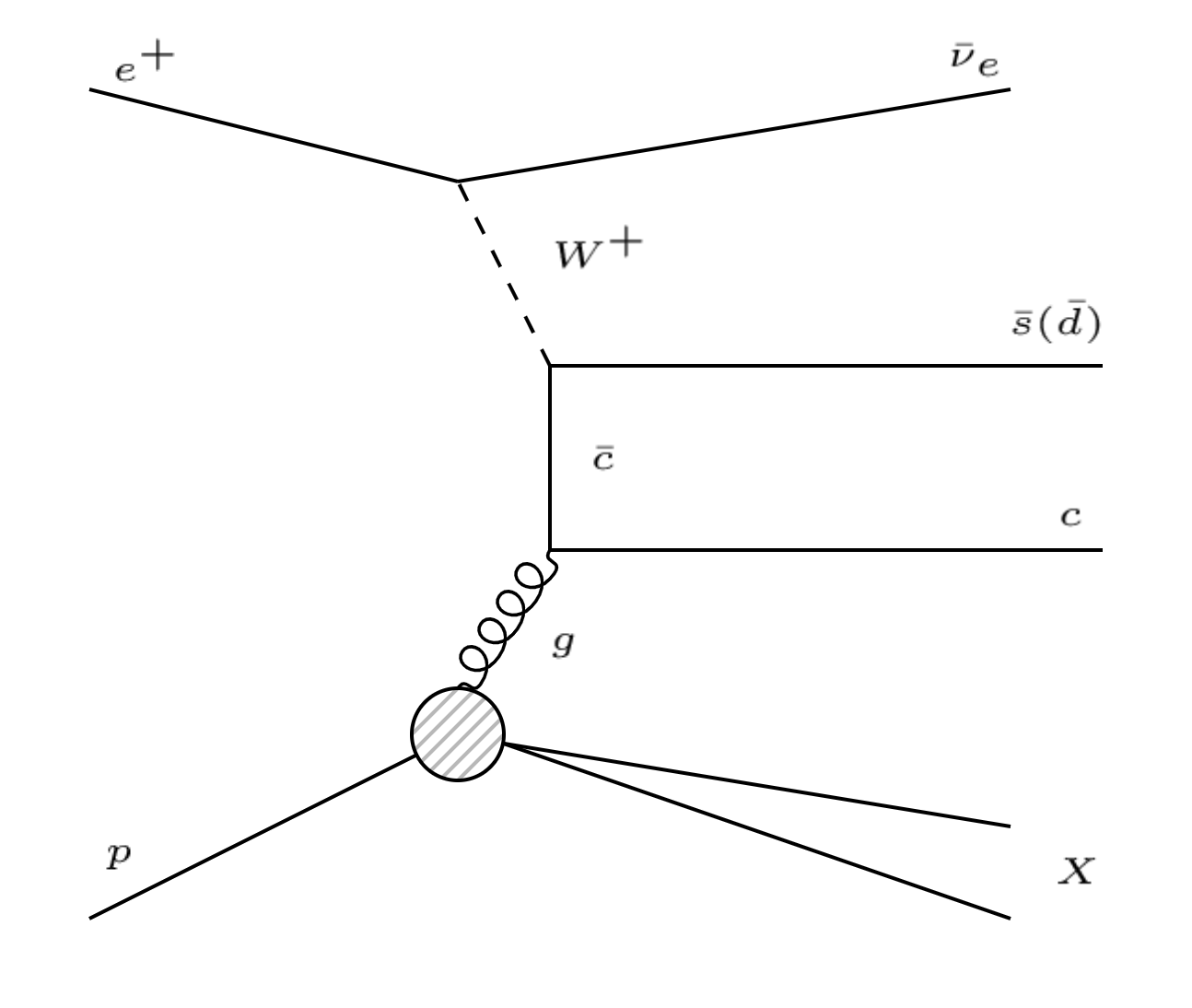}}
\label{fig:subim4}
\end{subfigure}
\caption{
  Feynman diagrams of charm-production subprocesses in $e^{+}p$ collisions.
  The QPM process illustrated in (i) describes $s(d) \rightarrow c$ transitions.
  In the QPM process (ii) $\bar{c} \rightarrow \bar{s}(\bar{d})$, the charm in the final state arises from the associated charm quark in the proton remnant X.
  In the BGF processes, the incoming $W$ boson couples to (iii) an $s\bar{s}(d\bar{d})$ or
  (iv) a $c\bar{c}$ pair from the gluon in the proton, producing a $c\bar{s}$ pair in the final state. 
}
\label{fig-subprocess1}
\vfill
\end{figure}

\begin{figure}[p]
\vfill
\centering
\includegraphics[width=8cm,height=6.7cm]{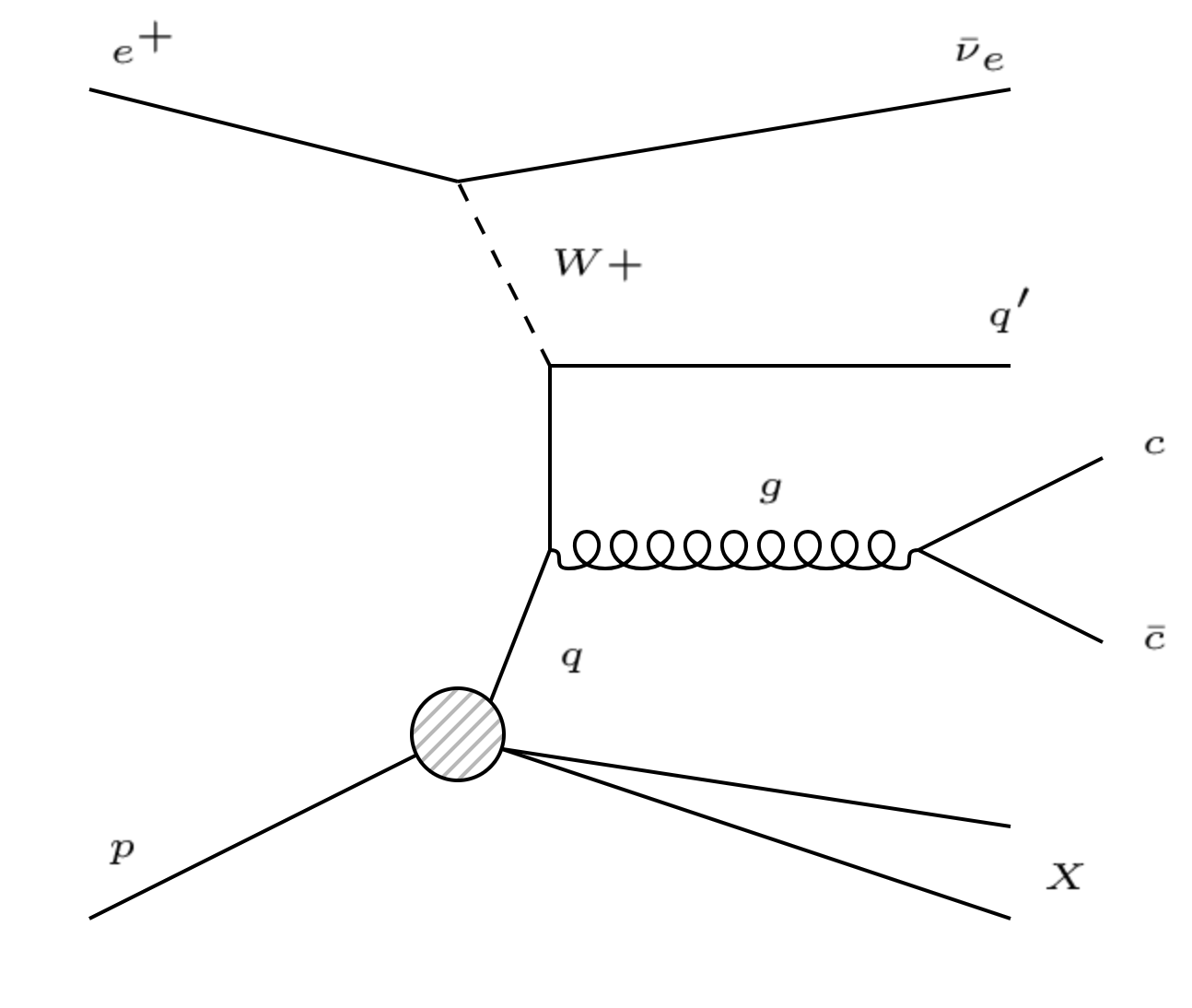} 
\label{fig:subim1}
\caption{
  Example Feynman diagram of QCD charm process. The $c\bar{c}$ pairs from the final-state gluons, illustrated in the figure, are referred to as QCD charm in the text.
}
\label{fig-subprocess2}
\vfill
\end{figure}

The differential cross section of charm production in CC DIS at HERA, mediated by a $W$ boson,
can be expressed in terms of the proton structure functions $F_{2}$, $xF_{3}$ and $F_{L}$ as follows~\cite{devenish:2003:dis}

\begin{equation}
\begin{aligned}  
  \frac{d^{2}\sigma(e^{\pm}p \rightarrow \bar{\nu}_{e}(\nu_{e})W^{\pm}X)}{dx_\textrm{Bj}dQ^{2}} = {} &
  \frac{G_{F}^{2}}{4\pi x_\textrm{Bj}}\frac{M^{4}_{W}}{(Q^{2}+M^{2}_{W})^{2}}[Y_{+}F_2(x_\textrm{Bj},Q^{2}) \mp Y_{-}xF_{3}(x_\textrm{Bj},Q^{2}) \\
 &\    - y^{2}F_{L}(x_\textrm{Bj},Q^{2})],
\end{aligned}
\end{equation}

where $G_{F}$ is the Fermi coupling constant, $M_{W}$ is the mass of the $W$ boson and $Y_{\pm} = 1 \pm (1-y)^2$.
The contribution from the longitudinal structure function, $F_{L}$, vanishes except at values of $y \approx 1$.
The basic electroweak single-charm production mechanisms have been outlined in Section~\ref{sec-int}.
In the leading-order plus parton-shower Monte Carlo (MC) simulation, the core electroweak matrix elements are based on the QPM graphs
in Fig.~\ref{fig-subprocess1} (i, ii) and BGF-like configurations in Fig.~\ref{fig-subprocess1} (iii, iv)  through initial-state
parton showering.
In addition, other tree-level higher-order processes are also added through leading-log (LL) parton showering.
The electroweak matrix elements involving only light quarks are complemented by occasional final-state gluon splitting into $c\bar{c}$
pairs in the parton shower, as depicted in Fig.~\ref{fig-subprocess2}, with a cutoff mimicking charm-mass effects.
At the single-event level, if only one of the two charm quarks (or its resulting hadron) is detected and its charge is not measured
(such as in the measurement technique used in this paper), then the contribution of this final-state QCD radiation is experimentally
indistinguishable from electroweak production.
The experimental measurement thus refers to a sum of all these processes, which make differing contributions to different regions
of phase space,
but cannot be disentangled with the presently available statistics.

In fixed-order QCD calculations, the final-state gluon-splitting contribution in Fig.~\ref{fig-subprocess2} is formally of
next-to-next-to-leading order (NNLO, $O(\alpha_{s}^{2})$) and thus not included in the next-to-leading-order (NLO, $O(\alpha_{s})$)
QCD predictions considered in this work, even though its contribution can be substantial.
Contributions from QPM-like (Fig.~\ref{fig-subprocess1} (i, ii)) and BGF-like (Fig.~\ref{fig-subprocess1} (iii, iv)) processes are separated
by the virtuality of the quark entering the electroweak process in relation to the chosen factorisation scale.
The NLO corrections to Fig.~\ref{fig-subprocess1} (i, ii) arise in the form of initial- or final-state gluon radiation, or a vertex correction.

In the zero-mass variable-flavour-number scheme (ZM-VFNS)~\cite{Thorne:2000zd, CooperSarkar:1997jk}, the charm part of the structure functions
$F^{c}_{2}$ and $xF^{c}_{3}$ can be expressed in terms of different PDFs as follows

\begin{align}
  F^{c}_{2} &= 2x_\textrm{Bj} \Big\{C_{2,q} \otimes \Big[|V_{cd}|^{2} \big(d + \bar{c}\big) + |V_{cs}|^{2} \big(s + \bar{c}\big)\Big] +
  2 \big(|V_{cd}|^{2} + |V_{cs}|^{2}\big) C_{2,g} \otimes g\Big\},\\
  xF^{c}_{3} &= 2x_\textrm{Bj} \Big\{C_{3,q} \otimes \Big[|V_{cd}|^{2} \big(d - \bar{c}\big) + |V_{cs}|^{2} \big(s - \bar{c}\big)\Big] +
  \big(|V_{cd}|^{2} + |V_{cs}|^{2}\big) C_{3,g} \otimes g\Big\},
\end{align}
in $e^{+}p$ collisions, and

\begin{align}
  F^{c}_{2} &= 2x_\textrm{Bj} \Big\{C_{2,q} \otimes \Big[|V_{cd}|^{2} \big(\bar{d} + c\big) + |V_{cs}|^{2} \big(\bar{s} + c\big)\Big] +
  2 \big(|V_{cd}|^{2} + |V_{cs}|^{2}\big) C_{2,g} \otimes g\Big\},\\
  xF^{c}_{3} &= 2x_\textrm{Bj} \Big\{C_{3,q} \otimes \Big[|V_{cd}|^{2} \big(-\bar{d} + c\big) + |V_{cs}|^{2} \big(-\bar{s} + c\big)\Big]+
  \big(|V_{cd}|^{2} + |V_{cs}|^{2}\big) C_{3,g} \otimes g\Big\},
\end{align}

in $e^{-}p$ collisions. Here $C_{i,j}$ is the coefficient function for parton $j$ in structure-function $F_i$ and $d$, $s$, $c$ and $g$
are respectively the down, strange, charm and gluon PDFs with the argument ($x_\textrm{Bj},Q^{2}$) omitted.
The parameters $|V_{ij}|$ are the Cabbibo--Kobayashi--Maskawa  matrix elements.
Part of the effects beyond NLO are resummed at next-to-leading log in the zero-mass approximation in this scheme. 

In the NLO fixed-flavour-number (FFN) scheme~\cite{Gluck:1996ve, Buza:1997mg}, charm-mass effects are treated explicitly up to
$O(\alpha_{s})$ in the matrix elements. In this scheme, there is no charm-quark content in the proton, thus the charm QPM graph in
Fig.~\ref{fig-subprocess1} (ii) and its associated higher-order corrections do not occur. This is compensated by a correspondingly larger
gluon content in the proton, such that all initial-state charm contributions irrespective of scale are treated explicitly in the BGF
matrix element (Fig.~\ref{fig-subprocess1} (iv)). No resummation is performed.

In the FONLL-B scheme~\cite{Cacciari:1998it, Forte:2010ta}, a general-mass variable-flavour-number scheme, charm-mass effects are accounted
for by interpolating between the ZM-VFNS and FFN predictions, such that all mass effects are correctly included up to $O(\alpha_s)$.

The xFitter framework~\cite{xfitter} was used to interface the theoretical predictions. 
Predictions in the FFN scheme were obtained from OPENQCDRAD~\cite{openqcd} using the ABMP 16.3 NLO PDF sets~\cite{abmp1, abmp2}.
Predictions in the FONLL-B scheme were obtained from APFEL~\cite{apfel} with NNPDF3.1~\cite{Ball:2017nwa}. 
The total uncertainties of the FFN and FONLL-B schemes were obtained by adding in quadrature the PDF, scale and charm-mass uncertainties.

In order to study the effects of strangeness suppression,
the ZM-VFNS predictions were obtained from QCDNUM~\cite{qcdnum} with HERAPDF2.0~\cite{Abramowicz:2015mha}.
The strange-quark fraction, $f_s = \bar{s}/(\bar{d}+\bar{s})$, was chosen to vary in the range
between a suppressed strange sea~\cite{Martin:2009iq,Nadolsky:2008zw} and an unsuppressed strange sea~\cite{atlas, atlasstrange}.
In addition, two more variations of the assumptions about the strange sea were made. Instead of assuming that the strange
contribution is a fixed fraction of the $d$-type sea, an $x$-dependent shape,  $x\bar{s}= 0.5 f_s' \tanh(-20(x-0.07))\, x\bar{D}$,
where $x\bar{D} = x\bar{d} + x\bar{s}$, was used in which high-$x$ strangeness is highly suppressed. This shape was suggested by
HERMES measurements~\cite{HERMES1,HERMES2}. The value of  $f_s'$ was also varied between $f_s'=0.3$ and $f_s'=0.5$.
The ZM-VFNS prediction was also evaluated with the ATLAS-$epWZ16$ PDF sets~\cite{atlas}.
 
\section{Experimental set-up}
\label{sec-exp}

This analysis was performed with data taken during the HERA II data-taking period in the years 2003--2007.
During this period, electrons and positrons with an energy of $27.5 \gev$ collided with protons with an energy of $920 \gev$ at a
centre-of-mass energy of $\sqrt{s} = 318 \gev$. The corresponding integrated luminosities are $173 \pb^{-1}$ and $185 \pb^{-1}$ for
$e^{+}p$ and $e^{-}p$ collisions, respectively.

\Zdetdesc

\Zctdmvdsttdesc{\ZcoosysfnBEeta}

\Zcaldesc

\Zbacdesc

\Zlumidesc
The fractional systematic
uncertainty on the measured luminosity was $2{\%}$.

\section{Monte Carlo simulation}
\label{sec-mc}

Inclusive CC DIS MC samples were generated to simulate the charm signal and the light-flavour (LF) background.
Neutral current DIS and photoproduction samples were used to simulate non-CC DIS backgrounds, which were
found to be negligible after the CC selection defined below.
The charged current events were generated with DJANGOH 1.6~\cite{django}, using the CTEQ5D PDF sets~\cite{cteq}
including QED and QCD radiative effects at the parton level. The ARIADNE 4.12 colour-dipole model~\cite{ariadne}
was used for parton showering. The Lund string model was used for hadronisation, as implemented
in JETSET 7.4.1~\cite{jetset}. The NC DIS events and photoproduction events were simulated by using DJANGOH
and HERWIG 5.9~\cite{herwig}, respectively.

\section{Event selection and reconstruction}
\label{sec-sel}

\subsection{Reconstruction of kinematic variables}
\label{sec-reco}

Charged current DIS at HERA produces a neutrino in the final state. The neutrino then escapes the ZEUS detector, resulting in a lack of
information on the leptonic final state. 
Thus, the Lorentz-invariant kinematic variables must be defined with the hadronic final state. In the present analysis,
this is done with the Jacquet--Blondel method, which assumes
the four-momentum of the exchange-boson $q$ to be equal not only to the difference in leptonic four-momentum $k-k'$
but also to that in hadronic four-momentum $p-p'$. Then, the invariant variables described in
Section~\ref{sec-charm-cc} can be reconstructed as

\begin{align}
y_{\textrm{JB}}         &= \frac{\sum_{h}(E-p_z)_h}{2 E_{e,\textrm{beam}}},\\
Q^{2}_{\textrm{JB}} &= \frac{p_{T,h}^{2}}{1-y_{\textrm{JB}}}, \\
x_{\textrm{JB}}        &= \frac{Q^{2}_{\textrm{JB}}}{sy_{\textrm{JB}}},
\end{align}

where
$E_{e,\textrm{beam}}$ is the electron beam energy, 
$\sum_{h}(E-p_{z})_h = \sum_i{(E_i-p_{{z},i})}$
is the hadronic $E-P_{z}$ variable with the sum extending over
the energies, $E_i$, and the longitudinal components of the momentum, $p_{{z},i}$
of the reconstructed hadronic final-state particles, $i$.
The quantity $p_{T,h} = \left| \sum_i {p}_{T,i} \right|$
is the total transverse momentum of the hadronic final state
with ${p}_{T,i}$
being the transverse-momentum vector of the particle $i$.
The mean value of the difference between the true and reconstructed kinematic variables
was found to be within $\approx 1\%$ in the MC simulation study.

\subsection{CC DIS selection}
The ZEUS online three-level trigger system loosely selected CC DIS candidates based on calorimeter and tracking
information~\cite{trigger2, trigger3}. The triggered events were then required to pass the following offline selection criteria
to reject non-CC DIS events:

\begin{itemize}
\item
  a kinematic selection cut was implemented at $200 \gev^{2} < Q^{2}_{\textrm{JB}} < 60000 \gev^{2}$ and $y_{\textrm{JB}} < 0.9$ to confine the sample
  into a region with good resolution of the kinematic quantities and small background;
\item 
  a characteristic of CC DIS events is the large missing transverse momentum, $p_{T,\textrm{miss}}$, in the calorimeter due to the undetected
  final-state neutrino. Events were required to have $p_{T,\textrm{miss}} > 12 \gev$ and $p'_{T,\textrm{miss}} > 10 \gev$, where $p'_{T,\textrm{miss}}$
  is the missing transverse momentum, excluding measurements taken from the CAL cells adjacent to the forward beam hole;
\item 
  further background rejection is discussed in detail in a dedicated study of CC DIS at ZEUS in the $e^{+}p$ scattering periods~\cite{CCzeus}.
  In addition, the remaining cosmic muons were removed by requiring the number of fired calorimeter cells $N_{\textrm{cell}} > 40$ and
  comparing fractions of energy deposited in the EMC and HAC. Events with energy deposited in the RCAL, $E_{\textrm{RCAL}} > 2 \gev$,
  were rejected if $E_{\textrm{RHAC}}/E_{\textrm{RCAL}} > 0.5$. 
  Events with energy in the BCAL, $E_{\textrm{BCAL}} > 2 \gev$, were rejected if $E_{\textrm{BHAC}} / E_{\textrm{BCAL}} > 0.85$,
  $E_{\textrm{BHAC1}} / E_{\textrm{BCAL}} > 0.7$ or $E_{\textrm{BHAC2}} / E_{\textrm{BCAL}} > 0.4$.
  Events with energy in the FCAL, $E_{\textrm{FCAL}} > 2 \gev$, were rejected if $E_{\textrm{FHAC}} / E_{\textrm{FCAL}} < 0.1$,
  $E_{\textrm{FHAC}} / E_{\textrm{FCAL}} > 0.85$, $E_{\textrm{FHAC1}} / E_{\textrm{FCAL}} > 0.7$ or $E_{\textrm{FHAC2}} / E_{\textrm{FCAL}} > 0.6$.

\end{itemize}

\begin{figure}[p]
\vfill
\begin{subfigure}{0.5\textwidth}
\phantomcaption
\stackinset{l}{1.7cm}{b}{6.3cm}{(\thesubfigure)}{\includegraphics[width=8cm,height=8cm]{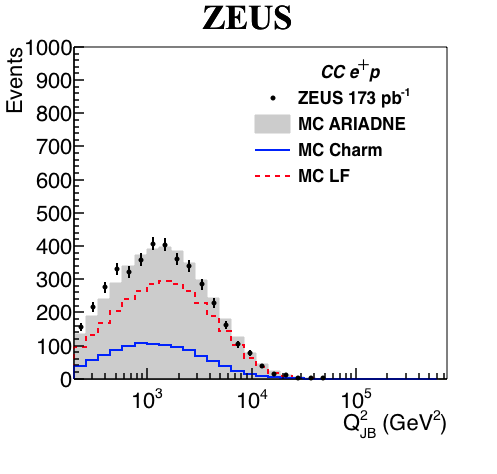}}
\label{fig:test}
\end{subfigure}
\begin{subfigure}{0.5\textwidth}
\phantomcaption
\stackinset{l}{1.7cm}{b}{6.3cm}{(\thesubfigure)}{\includegraphics[width=8cm,height=8cm]{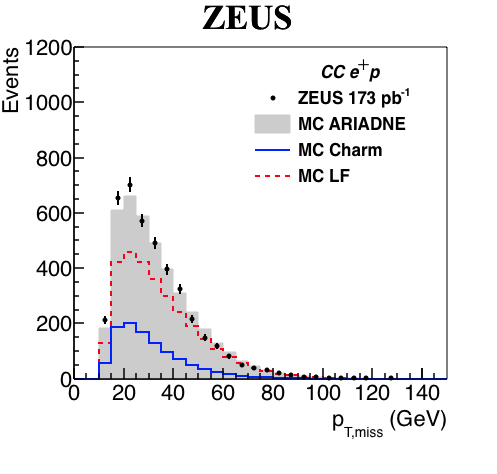}}
\label{fig:test}
\end{subfigure}
\begin{subfigure}{0.5\textwidth}
\phantomcaption
\stackinset{l}{1.7cm}{b}{6.3cm}{(\thesubfigure)}{\includegraphics[width=8cm,height=8cm]{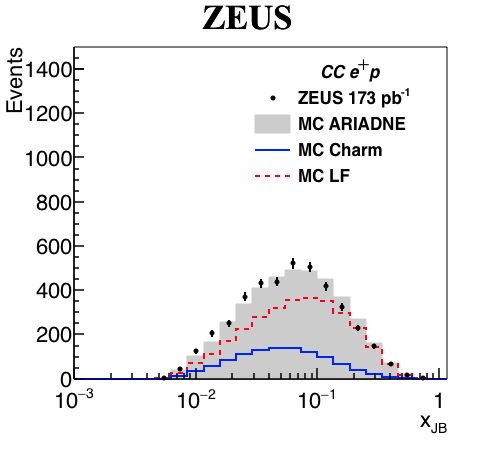}}
\label{fig:test}
\end{subfigure}
\begin{subfigure}{0.5\textwidth}
\phantomcaption
\stackinset{l}{1.7cm}{b}{6.3cm}{(\thesubfigure)}{\includegraphics[width=8cm,height=8cm]{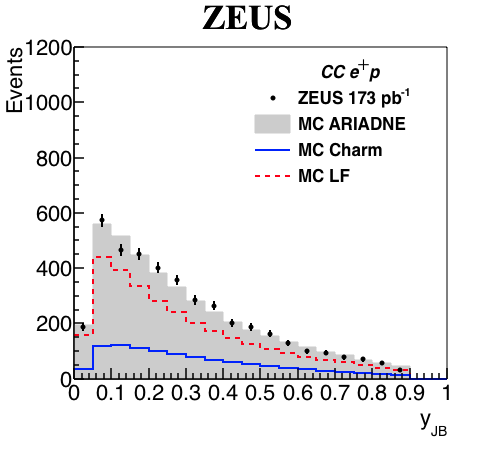}}
\label{fig:test}
\end{subfigure}

\caption{
  Comparison between data (dots) and MC (histogram) in kinematic variables (a) $Q^{2}_{\textrm{JB}}$, (b) $p_{T,\textrm{miss}}$, (c) $x_{\textrm{JB}}$ and (d) $y_{\textrm{JB}}$ for $e^{+}p$ collisions.
  The vertical error bars represent the statistical uncertainty in the data. "MC Charm" represents events with charm or anticharm quarks involved in the hard CC reaction either
  in the initial or final state. "MC LF" represents the contribution from light-flavoured events, i.e.\ with no heavy-flavour particles occurring in the event.
 }
\label{fig-event1}
\vfill
\end{figure}

\begin{figure}[p]
\vfill
\begin{subfigure}{0.5\textwidth}
\phantomcaption
\stackinset{l}{1.7cm}{b}{6.3cm}{(\thesubfigure)}{\includegraphics[width=8cm,height=8cm]{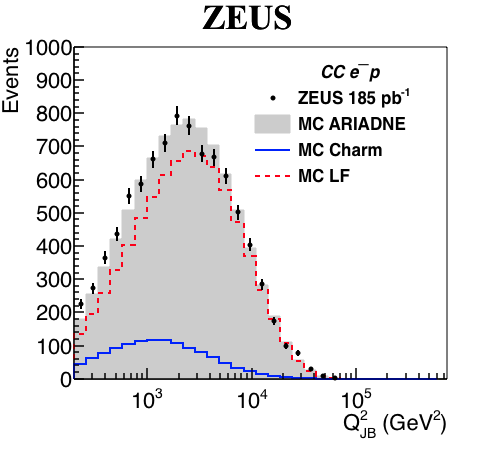}}
\label{fig:test}
\end{subfigure}
\begin{subfigure}{0.5\textwidth}
\phantomcaption
\stackinset{l}{2.7cm}{b}{6.3cm}{(\thesubfigure)}{\includegraphics[width=8cm,height=8cm]{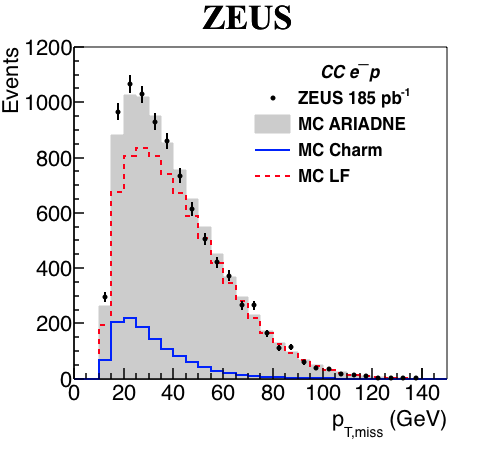}}
\label{fig:test}
\end{subfigure}
\begin{subfigure}{0.5\textwidth}
\phantomcaption
\stackinset{l}{6cm}{b}{6.3cm}{(\thesubfigure)}{\includegraphics[width=8cm,height=8cm]{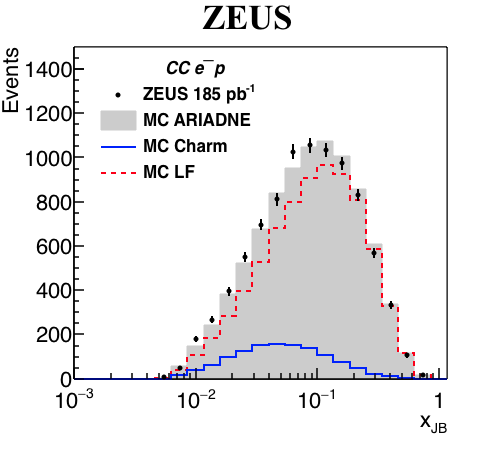}}
\label{fig:test}
\end{subfigure}
\begin{subfigure}{0.5\textwidth}
\phantomcaption
\stackinset{l}{2.5cm}{b}{6.3cm}{(\thesubfigure)}{\includegraphics[width=8cm,height=8cm]{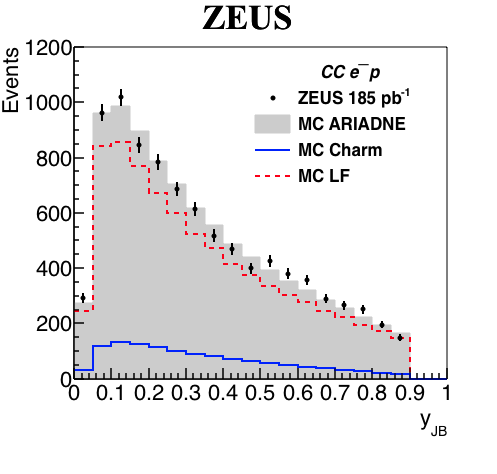}}
\label{fig:test}
\end{subfigure}
\caption{
  Comparison between data (dots)and MC (histogram) in kinematic variables (a) $Q^{2}_{\textrm{JB}}$, (b) $p_{T,\textrm{miss}}$, (c) $x_{\textrm{JB}}$ and (d) $y_{\textrm{JB}}$ for $e^{-}p$ collisions.
  The vertical error bars represent the statistical uncertainty in the data. "MC Charm" represents events with charm or anticharm quarks involved in the hard CC reaction either
  in the initial or final state. "MC LF" represents the contribution from light-flavoured events, i.e.\ with no heavy-flavour particles occurring in the event.
 }
\label{fig-event2}
\vfill
\end{figure}

A total of 4093 events in $e^{+}p$ data and 8895 events in $e^{-}p$ data passed these selection criteria.
Comparisons of data and MC at the event-level selection stage are shown in Figs.~\ref{fig-event1} and~\ref{fig-event2} for $e^+p$ and $e^-p$, respectively.
The MC distribution is consistent with the data in both the $e^{+}p$ and $e^{-}p$ periods.
From MC studies, the charm contribution to the CC events is expected to be about $25\%$ in the $e^{+}p$ periods and $12\%$ in the $e^{-}p$ periods and similar for both periods in terms of numbers of events.

\subsection{Charm selection and signal extraction}
\label{sec-charmsel}

Charm quarks in CC DIS events were tagged by using an inclusive lifetime method~\cite{zeus-cb, lifetime}.
In CC DIS at HERA, LF production has the highest production rate and is the major source of background.
The lifetime method uses the measurement of the decay length of the heavy-flavour (HF) particle to discriminate between signal and
background contributions.
The underlying principle of this method~\cite{zeus-cb} is that ground-state HF particles 
travel on average a measurable distance before they decay at a secondary vertex. 

Jets were reconstructed from energy-flow objects~\cite{epj:c1:81, thesis:briskin:1998},
which combine the information from calorimetry and tracking, corrected for energy loss
in the detector material. The $k_{T}$ clustering algorithm~\cite{Ellis:1993tq} was used
with a radius parameter $R = 1$ in the longitudinally invariant mode~\cite{Catani:1992zp, Catani:1993hr}.
The E-recombination scheme, which produces massive jets whose four-momenta are the sum of the four-momenta of the clustered objects, was used.
Events were selected if they contained at least one jet with transverse energy, $E_{T}^{\textrm{jet}}$, greater than $5 \gev$
and within the jet pseudorapidity range $-2.5 < \eta^{\textrm{jet}} < 2.0\ (1.5)$.\footnote{
  The tracking efficiency and resolution in the forward region $\eta^{\textrm{jet}} > 1.5$ suffered in the 2005 $(e^{-}p)$ data-taking period as the STT
  was turned off during this time. Thus, the jets from this period were required to satisfy a tighter $\eta^{\textrm{jet}}$ upper
  limit $\eta^{\textrm{jet}} = 1.5$.}
These selection criteria constrained the kinematic phase-space region of this analysis, along with the kinematic selection criteria at
the event-level selection stage. 

\begin{figure}[p]
\vfill
\begin{subfigure}{0.5\textwidth}
\phantomcaption
\stackinset{l}{1.7cm}{b}{6.3cm}{(\thesubfigure)}{\includegraphics[width=8cm,height=8cm]{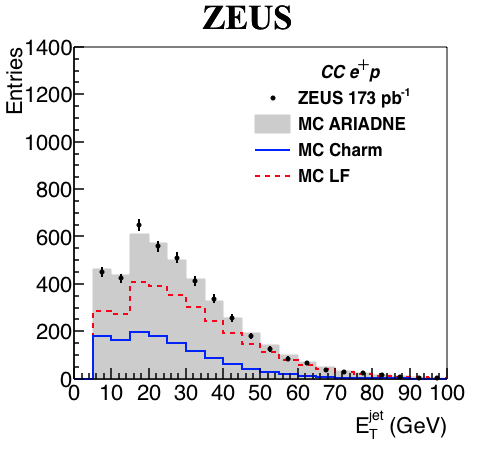}}
\label{fig:test}
\end{subfigure}
\begin{subfigure}{0.5\textwidth}
\phantomcaption
\stackinset{l}{1.7cm}{b}{6.3cm}{(\thesubfigure)}{\includegraphics[width=8cm,height=8cm]{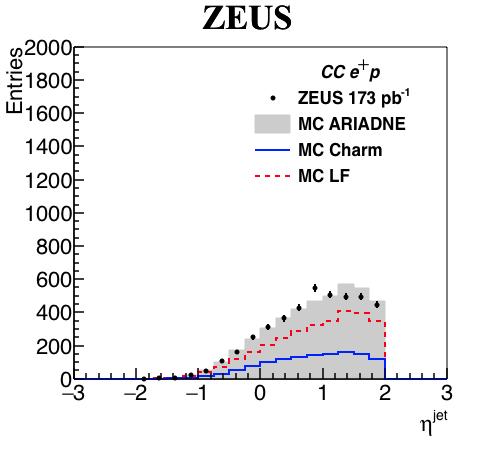}}
\label{fig:test}
\end{subfigure}
\begin{subfigure}{0.5\textwidth}
\phantomcaption
\stackinset{l}{1.7cm}{b}{6.3cm}{(\thesubfigure)}{\includegraphics[width=8cm,height=8cm]{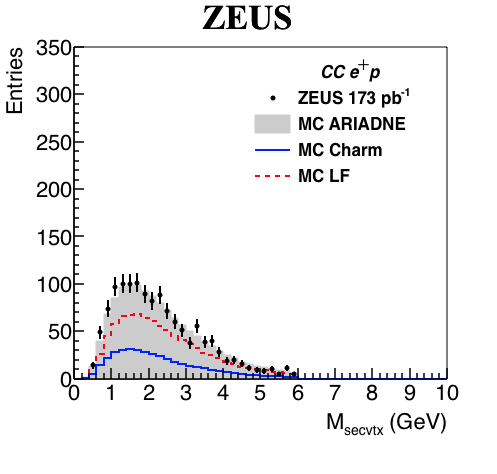}}
\label{fig:test}
\end{subfigure}
\begin{subfigure}{0.5\textwidth}
\phantomcaption
\stackinset{l}{1.7cm}{b}{6.3cm}{(\thesubfigure)}{\includegraphics[width=8cm,height=8cm]{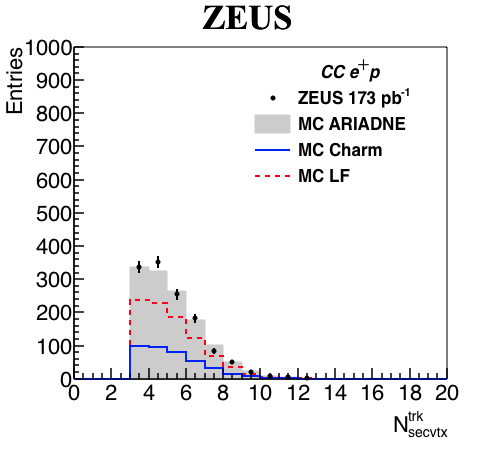}}
\label{fig:test}
\end{subfigure}

\caption{
  Comparison between data (points with vertical error bars) and MC (histogram) for jet and secondary-vertex distributions:
  (a) $E_{T}^{\textrm{jet}}$, (b) $\eta^{\textrm{jet}}$, (c) $M_{\textrm{secvtx}}$ and (d) $N^{\textrm{trk}}_{\textrm{secvtx}}$ for $e^{+}p$ collisions.
  The labels are the same as in Figs.~\ref{fig-event1} and~\ref{fig-event2}}
\label{fig-jet1}
\vfill
\end{figure}

\begin{figure}[p]
\vfill
\begin{subfigure}{0.5\textwidth}
\phantomcaption
\stackinset{l}{1.7cm}{b}{6.3cm}{(\thesubfigure)}{\includegraphics[width=8cm,height=8cm]{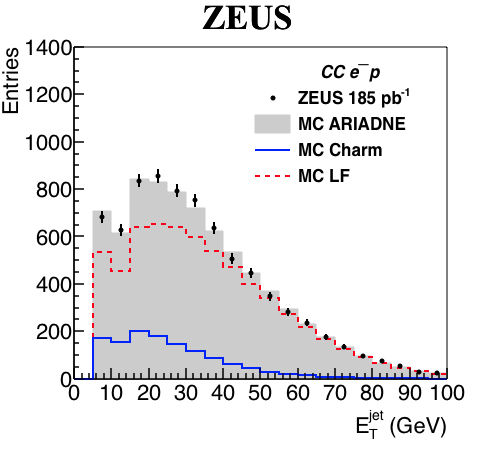}}
\label{fig:test}
\end{subfigure}
\begin{subfigure}{0.5\textwidth}
\phantomcaption
\stackinset{l}{6cm}{b}{6.3cm}{(\thesubfigure)}{\includegraphics[width=8cm,height=8cm]{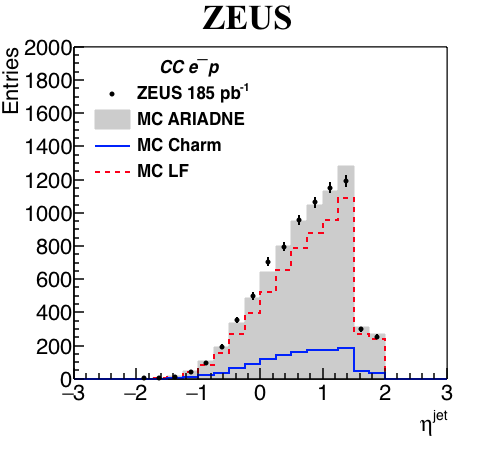}}
\label{fig:test}
\end{subfigure}
\begin{subfigure}{0.5\textwidth}
\phantomcaption
\stackinset{l}{1.7cm}{b}{6.3cm}{(\thesubfigure)}{\includegraphics[width=8cm,height=8cm]{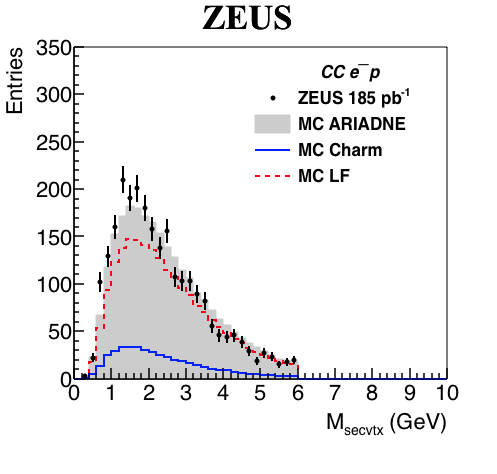}}
\label{fig:test}
\end{subfigure}
\begin{subfigure}{0.5\textwidth}
\phantomcaption
\stackinset{l}{1.7cm}{b}{6.3cm}{(\thesubfigure)}{\includegraphics[width=8cm,height=8cm]{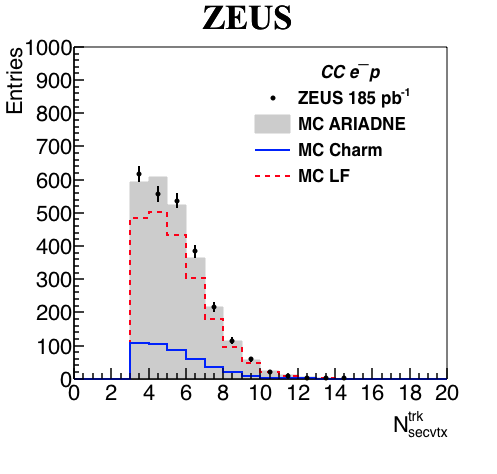}}
\label{fig:test}
\end{subfigure}

\caption{
  Comparison between data (points with vertical error bars) and MC (histogram) for jet and secondary-vertex distributions:
  (a) $E_{T}^{\textrm{jet}}$, (b) $\eta^{\textrm{jet}}$, (c) $M_{\textrm{secvtx}}$ and (d) $N^{\textrm{trk}}_{\textrm{secvtx}}$ for $e^{-}p$ collisions.
  The labels are the same as in Figs.~\ref{fig-event1} and~\ref{fig-event2}.}
\label{fig-jet2}
\vfill
\end{figure}

Tracks from the selected jets were required to have a transverse momentum, $p^{\textrm trk}_{T} > 0.5 \gev$, and the total number of hits in the MVD,
$N^{\textrm trk}_{MVD} \geq 4$ to reduce the effect of multiple scattering and ensure a good spatial resolution.
If more than two such tracks were associated with the jet, a secondary-vertex candidate was fitted from the selected tracks using
a deterministic annealing filter~\cite{PhysRevLett.65.945,Rose:1998dzq,DIDIERJEAN2010188}.
This fit provided the vertex position and its error matrix as well as the hadronic invariant mass, $M_{\textrm{secvtx}}$, of the charged tracks associated
with the reconstructed vertex. The charged-pion mass was assumed for all tracks when calculating the vertex mass.
The secondary-vertex candidates were required to satisfy the following criteria:

\begin{itemize}
\item
$N^{\textrm{trk}}_{\textrm{secvtx}} \geq 3$,
\item
$\chi^{2}/N_{dof} < 6$,
\item
$|z_{\textrm{secvtx}}| < 30\ \textrm{cm}$,
\item
$M_{\textrm{secvtx}} < 6 \gev$,
\item
$\sqrt{\Delta x^{2} + \Delta y^{2}} < 1\ \textrm{cm}$,
\end{itemize}

where $N^{\textrm{trk}}_{\textrm{secvtx}}$ is the number of tracks used to reconstruct the vertex, $\chi^{2}/N_{dof}$ is the goodness of the vertex
fitting, $z_{\textrm{secvtx}}$ is the $Z$-coordinate of the secondary vertex
and $\Delta x$, $\Delta y$ are the $X$- and $Y$-displacement of the secondary vertex from the primary interaction vertex.
These selection criteria ensure a good fit quality and high acceptance of the CTD and MVD for tracks used to reconstruct the vertices.
The requirement on the track multiplicity was implemented in order to reduce the number of background vertices.
Figures~\ref{fig-jet1} and~\ref{fig-jet2} show the distributions of the chosen jets and secondary-vertex candidates
for the $e^{+}p$ and $e^{-}p$ periods, respectively.

\begin{figure}[p]
\vfill
\begin{subfigure}{0.5\textwidth}
\phantomcaption
\stackinset{l}{6cm}{b}{6.3cm}{(\thesubfigure)}{\includegraphics[width=8cm,height=8cm]{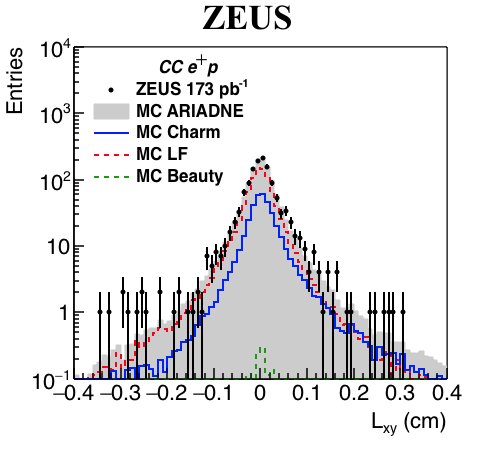}}
\label{fig:test}
\end{subfigure}
\begin{subfigure}{0.5\textwidth}
\phantomcaption
\stackinset{l}{6cm}{b}{6.3cm}{(\thesubfigure)}{\includegraphics[width=8cm,height=8cm]{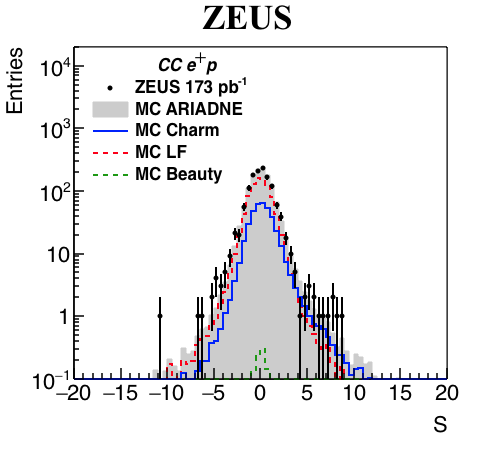}}
\label{fig:test}
\end{subfigure}
\begin{subfigure}{0.5\textwidth}
\phantomcaption
\stackinset{l}{1.7cm}{b}{6.3cm}{(\thesubfigure)}{\includegraphics[width=8cm,height=8cm]{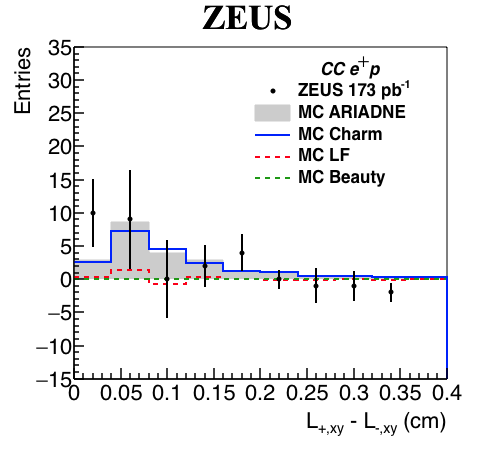}}
\label{fig:test}
\end{subfigure}
\begin{subfigure}{0.5\textwidth}
\phantomcaption
\stackinset{l}{2.5cm}{b}{6.3cm}{(\thesubfigure)}{\includegraphics[width=8cm,height=8cm]{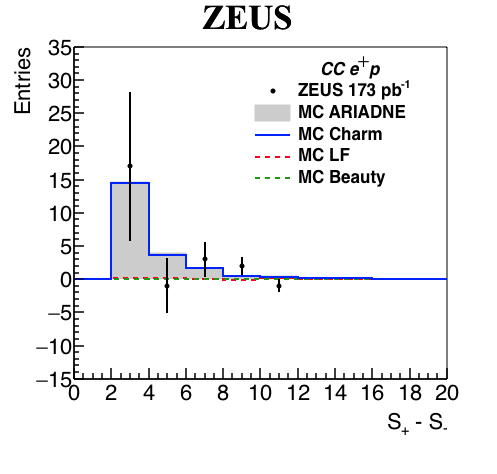}}
\label{fig:test}
\end{subfigure}
\caption{
  Comparison between data (points with vertical error bars) and MC (histogram) for $e^{+}p$ collisions for
  distributions of (a) the 2D decay length $L_{xy}$ and (b) significance $S$ distribution 
  and for distributions of the subtracted (c) decay-length $L_{+xy}-L_{-xy}$ and (d) significance $S_+-S_-$ distribution. The labels are the same as in Figs.~\ref{fig-event1} and~\ref{fig-event2}. "MC Beauty" represents events with beauty but no charm quark.
  }
\label{fig-decay1}
\vfill
\end{figure}

\begin{figure}[p]
\vfill
\begin{subfigure}{0.5\textwidth}
\phantomcaption
\stackinset{l}{6cm}{b}{6.3cm}{(\thesubfigure)}{\includegraphics[width=8cm,height=8cm]{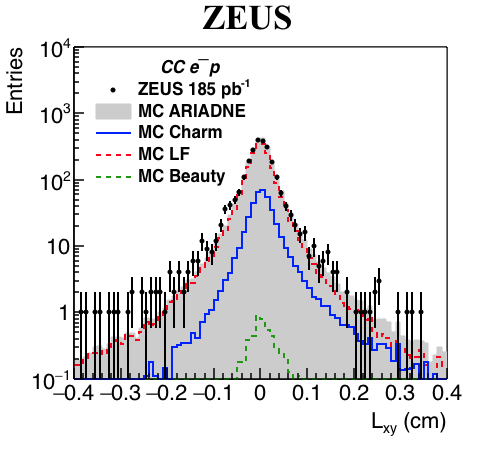}}
\label{fig:test}
\end{subfigure}
\begin{subfigure}{0.5\textwidth}
\phantomcaption
\stackinset{l}{6cm}{b}{6.3cm}{(\thesubfigure)}{\includegraphics[width=8cm,height=8cm]{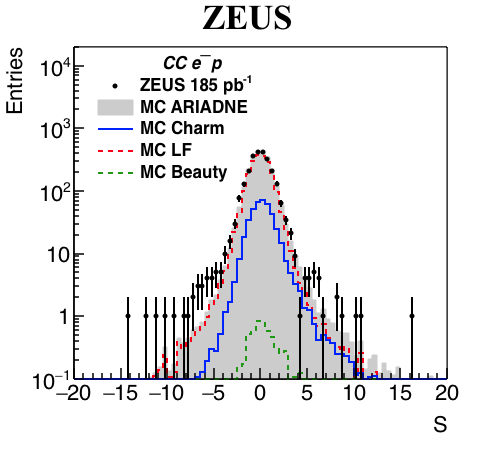}}
\label{fig:test}
\end{subfigure}
\begin{subfigure}{0.5\textwidth}
\phantomcaption
\stackinset{l}{1.7cm}{b}{6.3cm}{(\thesubfigure)}{\includegraphics[width=8cm,height=8cm]{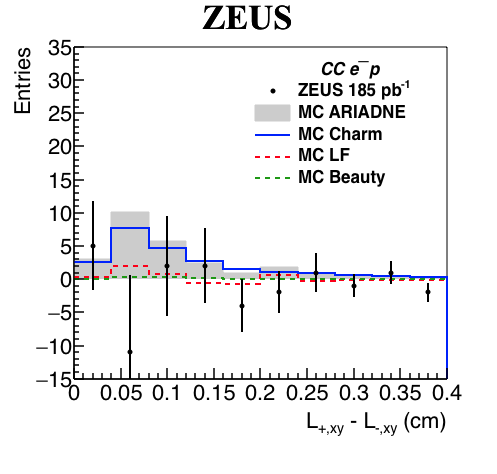}}
\label{fig:test}
\end{subfigure}
\begin{subfigure}{0.5\textwidth}
\phantomcaption
\stackinset{l}{1.7cm}{b}{6.3cm}{(\thesubfigure)}{\includegraphics[width=8cm,height=8cm]{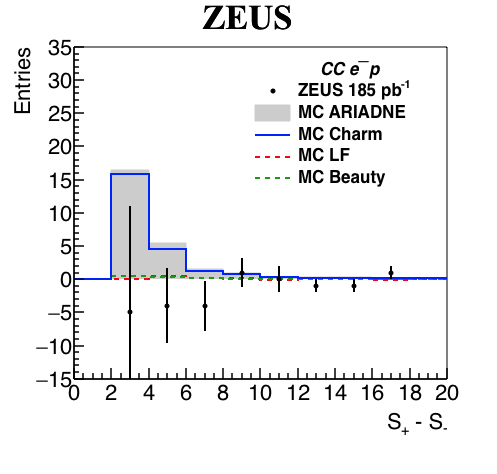}}
\label{fig:test}
\end{subfigure}
\caption{
  Comparison between data (points with vertical error bars) and MC (histogram) for $e^{-}p$ collisions for
  distributions of (a) the 2D decay length $L_{xy}$ and (b) significance $S$ distribution
    and for distributions of the subtracted (c) decay-length $L_{+xy}-L_{-xy}$ and (d) significance $S_+-S_-$ distribution. The labels are the same as in Figs.~\ref{fig-event1} and~\ref{fig-event2}. "MC Beauty" represents events with beauty but no charm quark.
  }
\label{fig-decay2}
\vfill
\end{figure}

The transverse decay length of the selected secondary vertices was projected onto the jet axis.
Due to the finite resolution of the MVD and the prompt production of LF particles, the distributions of the 2D decay length ($L_{xy}$) and the significance of the decay length ($S = L_{xy}/\delta_{L_{xy}}$) for LF jets were symmetric.
In contrast, the distributions for HF jets, in this case containing charmed particles, were asymmetric, as illustrated in Figs.~\ref{fig-decay1} and~\ref{fig-decay2} (a, b). 
A very small contribution from beauty is also shown; this is treated as background.
This enabled the LF background to be suppressed by subtracting the negative decay-length distribution from the positive decay-length distribution.

The region around $|L_{xy}| = 0$ or $|S| = 0$ is dominated by LF production, resulting in a large statistical uncertainty of the
distribution due to subtraction of two large numbers.
To optimise the precision of the extracted signal, vertex candidates were required to satisfy a significance threshold, $|S| > 2$.
Figures~\ref{fig-decay1} and~\ref{fig-decay2} (c, d) illustrate the shape of the variable distributions after the background subtraction.
The surviving events after the decay-length subtraction were used to extract charm cross sections in two bins of $Q^2$.

\section{Charm cross section}
\label{sec-cxs}

The lifetime method used in this analysis tags charm quarks regardless of their origin. Thus, the selected reactions include charm production
from final-state gluon splitting, such as shown in Fig.~\ref{fig-subprocess2}, which is here denoted by QCD charm, in addition to the
electroweak (EW) charm production discussed in Section 2.
In the present analysis, charm production was measured inclusively for  $200\gev^{2}<Q^{2}<60000\gev^{2}$ and $y<0.9$.
Additionally, to reflect the detector acceptance, a visible phase-space region was defined as: $200\gev^{2}<Q^{2}<60000\gev^{2}$,
$y<0.9$, $E_{T}^{\textrm{jet}}>5\gev$ and $-2.5<\eta^{\textrm{jet}}<2.0$. The limited statistics and absence of a charm-charge determination
prevented an experimental separation of the different theoretical contributions.
The visible charm-jet cross section, $\sigma_{c,\textrm{vis}}$, was initially measured as follows:

\begin{equation}
  \sigma_{c,\textrm{vis}} = \frac{N^{\textrm{data}} - N^{\textrm{MC}}_{\textrm{bg}}}{N^{\textrm{MC}}_{c}} \cdot \sigma^{\textrm{MC}}_{c,\textrm{vis}},
  \label{eq-vis}
\end{equation}

where $N^{\textrm{data}}$ is the reconstructed number of charm-jet candidates in the data after the $S_{+} - S_{-}$ subtraction,
$N^{\textrm{MC}}_{\textrm{bg}}$ is the background contribution and $N^{\textrm{MC}}_{c}$ is the charm/anti-charm contribution estimated from the MC.
Here $\sigma^{\textrm{MC}}_{c,\textrm{vis}}$ is the cross section of jets that are generated in the MC within the visible 
kinematic region and associated to a generated charm or anti-charm quark when $\sqrt{\Delta \phi^2 + \Delta \eta^2} < 1$,
where $\Delta \phi$ and $\Delta \eta$ are, respectively, the azimuthal angle and pseudorapidity difference between the jet and the charm quark.
Each charm quark was associated to the jet with the highest $E_{T}^{\textrm{jet}}$ satisfying the above criteria
and each such jet entered the visible cross section.
The different processes contributing to $\sigma^{\textrm{MC}}_{c,\textrm{vis}}$ as predicted by MC are given in Table~\ref{tab-contribution2}.

The EW contribution in the charm-quark signal, $\sigma^{\textrm{EW}}_{c,\textrm{vis}}$, should be evaluated by subtracting the QCD contribution from
gluon splitting (Fig.~\ref{fig-subprocess2}). However, the prediction from ARIADNE 4.12, like any 
prediction from gluon splitting in the massless mode with cutoff, cannot be considered to be reliable.
Since the contribution predicted by ARIADNE (see Table~\ref{tab-contribution2}) is both small and imprecise, it was not subtracted but
rather included in the systematic uncertainties. The visible jet cross section was extrapolated and converted to the total EW cross section
via a factor $C_{\textrm{ext}}$, calculated from the ratio of the number of charm events generated in the full kinematic range,
$N^{\textrm{EW}}_{\textrm{gen}}$, to the number of charm jets of EW origin within the visible kinematic region, $N^{\textrm{EW}}_{\textrm{vis}}$:

\begin{align}
C_{\textrm{ext}} &= \frac{N^{\textrm{EW}}_{\textrm{gen}}}{N^{\textrm{EW}}_{\textrm{vis}}}.
\end{align}
The resulting total EW charm cross section, $\sigma_{c^{\textrm{EW}}}$, is then given by

\begin{align}
\sigma_{c^{\textrm{EW}}} &= C_{\textrm{ext}} \ \sigma_{c, \textrm{vis}} \nonumber \\
&= \frac{N^{\textrm{EW}}_{\textrm{gen}}}{N^{\textrm{EW}}_{\textrm{vis}}} \ \frac{N^{\textrm{data}} - N^{\textrm{MC}}_{\textrm{bg}}}{N^{\textrm{MC}}_{c}} \ \sigma^{\textrm{MC}}_{c,\textrm{vis}}.
\label{eq-sigmaew}
\end{align}
This is predicted by the ARIADNE MC to be approximately $9~\pb$.

\section{Systematic uncertainties}
\label{sec-syst}

Although the statistical power of the current data is limited, it is important for future studies to understand the limitations
of the current method by careful evaluation of the systematic uncertainties. The sources
of uncertainty and their estimated effects on the total EW charm cross sections provided in parentheses
($\delta \sigma^{e^{+}p},\; \delta \sigma^{e^{-}p}$) are:

\begin{itemize}

\item{Secondary vertex rescaling}\\
  The MC samples used in this analysis produced a higher fraction of events with secondary vertices than the data.
  For the nominal result, $N^{\textrm{MC}}_{c}$ and $N^{\textrm{MC}}_{\textrm{bg}}$ in Eq.~\ref{eq-vis} were reduced proportionally.
  For the systematic uncertainty, only $N^{\textrm{MC}}_{\textrm{bg}}$ was rescaled ($-1.2 \pb,\; +0.9 \pb$).

\item{EW charm fraction}\\
  The MC predictions of the QCD contribution (Fig.~\ref{fig-subprocess2}) shown in Table~\ref{tab-contribution2} of $+6 \%$ for $e^{+}p$ collisions
  and $+12 \%$ for $e^{-}p$ collisions were taken as systematic uncertainty ($-0.6 \pb,\; - 1.1 \pb$). 

\item{LF background}\\
  The uncertainty due to the remaining LF background was estimated by varying it by $\pm 30 \%$~\cite{zeus-cb}
  ($\pm 0.1 \pb,\; \pm 0.3 \pb$).

\item{CC DIS  selection}\\
  The uncertainty due to the CC selection cuts was estimated by varying these cuts as in the previous ZEUS analysis~\cite{Chekanov:2003vw}
  ($\pm 0.2 \pb,\; \pm 0.1 \pb$).  

\item{Jet energy scale}\\
  The part of the transverse jet energy measured in the calorimeter in the MC was varied by its estimated uncertainty
  of $\pm 3\%$ ($\pm 0.0 \pb,\; \pm 0.1 \pb$).

\end{itemize}

These uncertainties were added in quadrature.
The uncertainty in the ZEUS luminosity measurement is $\pm 2 \%$ and was not included in the results.

In addition, the effect of the significance cut, $|S| > 2$, was studied.
Small changes in the value of the significance cut resulted in large changes of the extracted signal.
This was found to be due to statistical fluctuations in the number of events in the region close to the $|S|$ lower cut value.
From a dedicated study, the effects on the cross sections were found to be as large as $\pm 5$~pb.
As this result was still strongly affected by statistical fluctuations, which have been included in the quoted statistical uncertainty,
it was not included in the systematic uncertainty.

Additionally, the uncertainty in the secondary-vertex selection method was estimated by
reducing the requirement on the number of tracks, $N^{\textrm{trk}}_{\textrm{secvtx}}$, from three to two.
The effects on the cross sections were found to be as large as $+ 3$~pb.
This was again strongly affected by statistical fluctuations and not included in the systematic uncertainty.

\section{Results}
\label{sec-results}

\begin{figure}[p]
\vfill
\begin{subfigure}{0.5\textwidth}
\phantomcaption
\stackinset{l}{1.7cm}{b}{6.3cm}{(\thesubfigure)}{\includegraphics[width=8cm,height=8cm]{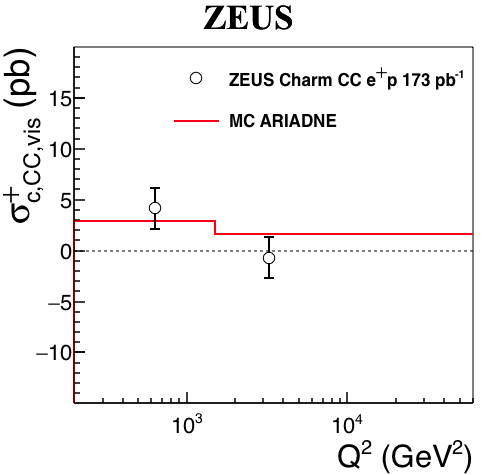}}
\label{fig:test}
\end{subfigure}
\begin{subfigure}{0.5\textwidth}
\phantomcaption
\stackinset{l}{1.7cm}{b}{6.3cm}{(\thesubfigure)}{\includegraphics[width=8cm,height=8cm]{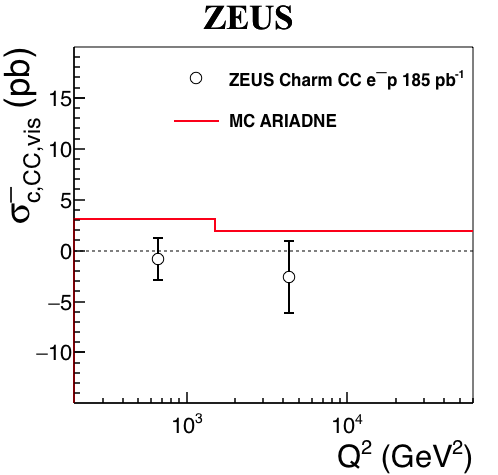}}
\end{subfigure}
\vfill
\caption{
  The visible charm cross sections, $\sigma_{c,\textrm{vis}}$, in two bins in $Q^{2}$ for (a) $e^{+}p$  and (b) $e^{-}p$ collisions.
  The vertical error bars show the total uncertainties; the systematic uncertainties are negligible. 
  The solid lines represent predictions obtained with the ARIADNE MC. 
}
\label{fig-summary1}
\end{figure}

The charm-jet cross sections in CC DIS in $e^{\pm}p$ collisions were measured in the visible kinematic phase space of
$200\gev^{2}<Q^{2}<60000\gev^{2}$, $y<0.9$, $E_{T}^{\textrm{jet}}>5\gev$ and $-2.5<\eta^{\textrm{jet}}<2.0$ to be

\begin{alignat*}{7}
&\sigma^{+}_{c,\textrm{vis}} &&= \ \  4.0&&\ \pm 2.8&&\ (\textrm{stat.})&&\ ^{+0.1}_{-0.6}&&\ (\textrm{syst.})&&\ \pb,\\
&\sigma^{-}_{c,\textrm{vis}} &&= -3.0&&\ \pm 3.8&&\ (\textrm{stat.})&&\ ^{+0.5}_{-0.1}&&\ (\textrm{syst.})&&\ \pb,
\end{alignat*}
where the superscript $\pm$ denotes the charge of the incoming lepton. In addition, the cross sections were obtained for
two separate $Q^2$ bins, $200\gev^{2}<Q^{2}<1500\gev^{2}$ and $1500\gev^{2}<Q^{2}<60000\gev^{2}$, and are shown in Fig.~\ref{fig-summary1}.

The total electroweak charm cross sections were found, following Eq.~\ref{eq-sigmaew}, to be

\begin{alignat*}{7}
&\sigma^{+}_{c^{\textrm{EW}}} &&= \ \   8.5&&\ \pm 5.5&&\ (\textrm{stat.})&&\ ^{+0.2}_{-1.3}&&\ (\textrm{syst.})&&\ \pb,\\
&\sigma^{-}_{c^{\textrm{EW}}} &&= -5.7&&\ \pm 7.2&&\ (\textrm{stat.})&&\ ^{+1.0}_{-1.2}&&\ (\textrm{syst.})&&\ \pb.
\end{alignat*}

\begin{figure}[p]
\vfill
\begin{subfigure}{0.5\textwidth}
\phantomcaption
\stackinset{l}{1.7cm}{b}{6.3cm}{(\thesubfigure)}{\includegraphics[width=8cm,height=8cm]{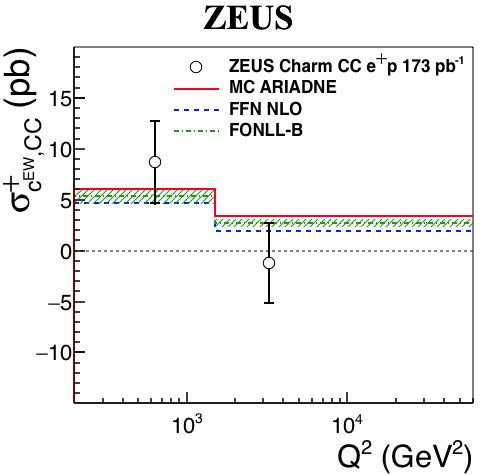}}
\label{fig:test}
\end{subfigure}
\begin{subfigure}{0.5\textwidth}
\phantomcaption
\stackinset{l}{1.7cm}{b}{6.3cm}{(\thesubfigure)}{\includegraphics[width=8cm,height=8cm]{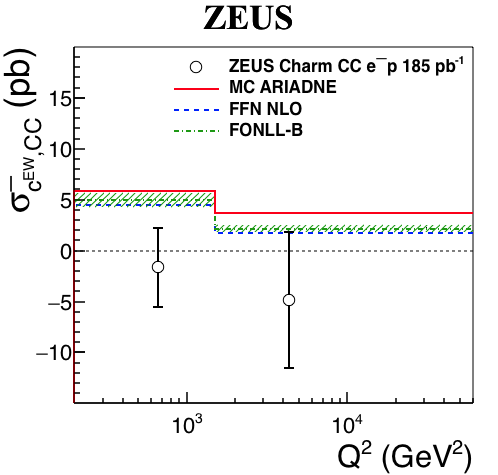}}
\label{fig:sigma2}
\end{subfigure}
\vfill
\caption{
  The EW charm cross sections, $\sigma_{c^{\textrm{EW}}}$, in two bins of $Q^{2}$ for (a) $e^{+}p$ and (b) $e^{-}p$ collisions.
  The vertical error bars show the total uncertainties; the included systematic uncertainties are negligible.
  The solid lines represent predictions obtained with the ARIADNE MC. The dashed and dashed-dotted lines represent,
  respectively, predictions from the FFN and FONLL-B schemes.
  Hatched bands are the total uncertainty in the predictions from FONLL-B schemes.
}
\label{fig-summary2}
\end{figure}

The QCD contribution to charm production was introduced as an additional systematic uncertainty.
Theory predictions obtained at NLO QCD with the FFN and FONLL-B schemes are compared to the data in bins of $Q^{2}$ in Fig.~\ref{fig-summary2}.
Table~\ref{tab-cross} provides the experimental values of the cross sections $\sigma_{c,\textrm{vis}}$ and $\sigma_{c^{\textrm{EW}}}$ for the two bins
in $Q^2$. The contributions of the charm production subprocesses to the final EW cross section in each bin were estimated in the ARIADNE MC,
FFN and FONLL-B predictions and are listed in Table~\ref{tab-contribution1}.  In Table~\ref{tab-prediction1}, the theory predictions from the FFN
and FONLL schemes are shown with the total uncertainties, as discussed in Section~\ref{sec-charm-cc}.
The predictions from the ZM-VFNS scheme with varied strange-quark fraction are given in Table~\ref{tab-prediction2}.
A further reduction of the theory uncertainties can be achieved in the future by including NNLO corrections~\cite{Blumlein:2014fqa}.

The theory predictions in Table~\ref{tab-contribution1} suggest that the most interesting subprocess, namely the QPM process depicted in
Fig.~\ref{fig-subprocess1} (i), contributes about $30-50 \%$ to the final EW cross section,
depending on the kinematic range and QCD scheme used. In general, the data are well described by the theory predictions,
however the large experimental uncertainties prevent a discrimination between the different models.

\section{Summary and outlook}
\label{sec-outlook}

Measurements of charm production in charged current deep inelastic scattering in $e^{\pm}p$ collisions have been performed
based on HERA II data with an integrated luminosity of $358 \pb^{-1}$, which corresponds to $e^{+}p$ collisions with an integrated
luminosity of $173 \pb^{-1}$ and $e^{-}p$ collisions with an integrated luminosity of $185 \pb^{-1}$. Visible charm-jet cross
sections for each lepton beam type were measured within a kinematic region
$200\gev^{2}<Q^{2}<60000\gev^{2}$, $y<0.9$, $E_{T}^{\textrm{jet}}>5\gev$ and $-2.5<\eta^{\textrm{jet}}<2.0$.
They were extrapolated to the EW cross sections given in the kinematic range $200\gev^{2}<Q^{2}<60000\gev^{2}$ and $y<0.9$.
Theoretical predictions with several assumptions about the 
strange-quark content of the proton and using different heavy-flavour schemes were found to be consistent with the data within
the large experimental uncertainties. The analysis presented here shows the potential of DIS measurements to increase the
knowledge about the strange-quark content of the proton. 
Future lepton--ion collider projects such as the electron--ion collider~\cite{Accardi:2012qut} or LHeC~\cite{lhec} will have much
higher luminosity than HERA, accompanied by improved vertex detection capabilities.
These projects should then be able to make an important contribution to the knowledge
of the strange-quark content of the proton.

\section*{Acknowledgements}
\label{sec-ack}

\Zacknowledge

\clearpage

{
\ifzeusbst
  \ifzmcite
     \bibliographystyle{./l4z_default3}
  \else
     \bibliographystyle{./l4z_default3_nomcite}
  \fi
\fi
\ifzdrftbst
  \ifzmcite
    \bibliographystyle{./l4z_draft3}
  \else
    \bibliographystyle{./l4z_draft3_nomcite}
  \fi
\fi
\ifzbstepj
  \ifzmcite
    \bibliographystyle{./l4z_epj3}
  \else
    \bibliographystyle{./l4z_epj3_nomcite}
  \fi
\fi
\ifzbstjhep
  \ifzmcite
    \bibliographystyle{./l4z_jhep3}
  \else
    \bibliographystyle{./l4z_jhep3_nomcite}
  \fi
\fi
\ifzbstnp
  \ifzmcite
    \bibliographystyle{./l4z_np3}
  \else
    \bibliographystyle{./l4z_np3_nomcite}
  \fi
\fi
\ifzbstpl
  \ifzmcite
    \bibliographystyle{./l4z_pl3}
  \else
    \bibliographystyle{./l4z_pl3_nomcite}
  \fi
\fi
{\raggedright
\bibliography{./syn.bib,%
              ./myref.bib,%
              ./l4z_zeus.bib,%
              ./l4z_h1.bib,%
              ./l4z_articles.bib,%
              ./l4z_books.bib,%
              ./l4z_conferences.bib,%
              ./l4z_misc.bib,%
              ./l4z_preprints.bib}}
}
\vfill\eject
\begin{table}[]
  \begin{center}
     \renewcommand{\arraystretch}{1.2}
\begin{tabular}{|c|cccc|}
\hline
    \multirow{2}{*}{$e^{+}p$}    & \multicolumn{4}{c|}{MC Contribution (\%)} \\
\cline{2-5}
      & $d \rightarrow c$    & $s \rightarrow c$  & $\bar{c} \rightarrow \bar{s}(\bar{d})$ & $g \rightarrow c\bar{c}$   \\
\hline
 $\sigma^{\textrm{MC}}_{c,\textrm{vis}} + \sigma(g \rightarrow c\bar{c})$  &    9     &    45     &    40     &    6     \\
 $\sigma^{\textrm{MC}}_{c^{\textrm{EW}}} + \sigma(g \rightarrow c\bar{c})$  &    7     &    31     &    58     &     4      \\
\hline
\end{tabular}
\\
\begin{tabular}{|c|cccc|}
\hline
     \multirow{2}{*}{$e^{-}p$}   & \multicolumn{4}{c|}{MC Contribution (\%)} \\
\cline{2-5}
     & $\bar{d} \rightarrow \bar{c}$    & $\bar{s} \rightarrow \bar{c}$  & $c \rightarrow s(d)$ & $g \rightarrow c\bar{c}$   \\
\hline
 $\sigma^{\textrm{MC}}_{c,\textrm{vis}} + \sigma(g \rightarrow c\bar{c})$ &    3     &   45      &  40       &  12       \\
 $\sigma^{\textrm{MC}}_{c^{\textrm{EW}}}+ \sigma(g \rightarrow c\bar{c})$ &      2   &      31   &   57      &      10     \\
\hline
\end{tabular}
\caption{
  MC contributions (\%) of charm subprocesses to $\sigma^{\textrm{MC}}_{c,\textrm{vis}}$ and $\sigma^{\textrm{MC}}_{c^{\textrm{EW}}}$ as
  predicted by ARIADNE.
  The first two columns ($d \rightarrow c$ and $s \rightarrow c$ for $e^{+}p$ collisions, for example) reflect the contributions
  from the QPM processes described in Fig.~\ref{fig-subprocess1} (i) and a higher-order correction described in
  Fig.~\ref{fig-subprocess1} (iii).
  The contribution of the final-state gluon splitting described in Fig.~\ref{fig-subprocess2} enters the fourth column
  ($g \rightarrow c\bar{c}$).
	}
  \label{tab-contribution2}
\end{center}
\end{table}

\begin{table}[p]
  \begin{center}
     \renewcommand{\arraystretch}{1.2}
\begin{tabular}{@{}|r@{--}l|rlrlr|rlrlr|}
\hline
\multicolumn{2}{|c|}{\begin{tabular}[x]{@{}c@{}}$Q^{2}$ range\\$(\gev^{2})$\end{tabular}}&\multicolumn{5}{c|}{$\sigma_{c, \textrm{vis}} (\pb)$}&\multicolumn{5}{c|}{$\sigma_{c^{\textrm{EW}}} (\pb)$}\\ 
\hline
\multicolumn{12}{|c|}{$e^{+}p$}\\
\hline
$200$ & $1500$  & $ 4.1 $ & $\pm 2.0$ & (stat.) & $^{+0.1}_{-0.6}$ & (syst.) & $ 8.7 $& $\pm 4.1$ & (stat.) & $^{+0.2}_{-1.4}$ & (syst.)\\
$1500$ & $60000$& $ -0.7 $& $\pm 2.0$ & (stat.) & $^{+0.2}_{-0.0}$ & (syst.) & $ -1.2 $& $\pm 3.9$ & (stat.) & $^{+0.3}_{-0.3}$ & (syst.)\\
\hline
\multicolumn{12}{|c|}{$e^{-}p$}\\
\hline
$200$ & $1500$  & $ -0.9 $& $\pm 2.1$ & (stat.) & $^{+0.2}_{-0.0}$ & (syst.) & $ -1.7 $& $\pm 3.9$ & (stat.) & $^{+0.3}_{-0.3}$ & (syst.)\\
$1500$ & $60000$& $ -2.6 $& $\pm 3.5$ & (stat.) & $^{+0.5}_{-0.1}$ & (syst.) & $ -4.8 $& $\pm 6.7$ & (stat.) & $^{+0.9}_{-0.8}$ & (syst.)\\
\hline
\end{tabular}
\caption{
Measured visible cross sections, $\sigma_{c, \textrm{vis}}$, and EW cross section, $\sigma_{c^{\textrm{EW}}}$, for two $Q^{2}$ bins.
	}
  \label{tab-cross}
\end{center}
\end{table}

\begin{table}[]
  \begin{center}
     \renewcommand{\arraystretch}{1.2}
\begin{tabular}{|c|c|c|ccc|ccc|}
\hline
\multicolumn{3}{|c|}{\multirow{3}{*}{$e^{+}p$}} & \multicolumn{6}{c|}{Contribution (\%)} \\
\cline{4-9}
\multicolumn{3}{|c|}{} & \multicolumn{3}{c|}{$200 < Q^2 < 1500 \gev^2$} & \multicolumn{3}{c|}{$1500 < Q^2 < 60000 \gev^2$} \\
\cline{4-9}
\multicolumn{3}{|c|}{} & $d \rightarrow c$ & $s \rightarrow c$ & $\bar{c} \rightarrow \bar{s}(\bar{d})$ & $d \rightarrow c$ & $s \rightarrow c$ & $\bar{c} \rightarrow \bar{s}(\bar{d})$         \\
\hline
\multicolumn{3}{|c|}{ARIADNE MC}                            & 6   & 36 & 58 &             10 & 26 & 64          \\
\multicolumn{3}{|c|}{FFN NLO ABMP16.3}                      & 8   & 49 & 43 &             16 & 43 & 41          \\
\multicolumn{3}{|c|}{FONLL-B NNPDF3.1}                      & 8   & 43 & 49 &             12 & 37 & 51          \\
\hline
\hline
\multicolumn{3}{|c|}{\multirow{3}{*}{$e^{-}p$}} & \multicolumn{6}{c|}{Contribution (\%)}  \\
\cline{4-9}
\multicolumn{3}{|c|}{} & \multicolumn{3}{c|}{$200 < Q^2 < 1500 \gev^2$} & \multicolumn{3}{c|}{$1500 < Q^2 < 60000 \gev^2$} \\
\cline{4-9}
\multicolumn{3}{|c|}{} &  $\bar{d} \rightarrow \bar{c} $  &  $\bar{s} \rightarrow \bar{c} $  &   $c \rightarrow s(d)$ &  $\bar{d} \rightarrow \bar{c} $  &  $\bar{s} \rightarrow \bar{c} $  &   $c \rightarrow s(d)$   \\    
\hline
\multicolumn{3}{|c|}{ARIADNE MC}                            & 3   & 37 & 60 &              2 & 29  & 69          \\
\multicolumn{3}{|c|}{FFN NLO ABMP16.3}                      & 4   & 51 & 45 &              5 & 49 & 46          \\
\multicolumn{3}{|c|}{FONLL-B NNPDF3.1}                      & 4   & 43 & 53 &             4 & 33 & 63          \\
\hline
\end{tabular}
\caption{
  Contribution (\%) of charm subprocesses to EW charm production in CC DIS in both $e^{+}p$ and $e^{-}p$ collisions,
  as predicted by the ARIADNE MC and FFN and FONLL-B schemes. The labels  are explained in Table~\ref{tab-contribution2}.
  Additionally for the MC and FONLL-B scheme, the contribution of the QPM process in Fig.~\ref{fig-subprocess1} (ii) enters
  in the third column ($\bar{c} \rightarrow \bar{s}(\bar{d})$) with a higher-order correction from the BGF process in
  Fig.~\ref{fig-subprocess1} (iv).
  For the FFN scheme, the process described in Fig.~\ref{fig-subprocess1} (ii) does not participate.
  Thus the content of the third column is provided by the BGF process of Fig.~\ref{fig-subprocess1} (iv) only.
}
\label{tab-contribution1}
\end{center}
\end{table}

\begin{table}[]
  \begin{center}
     \renewcommand{\arraystretch}{1.2}
\begin{tabular}{|c@{--}c|c|c|c|c|c|c|c|c|}
\hline
\multicolumn{2}{|c|}{\multirow{4}{*}{\begin{tabular}[x]{@{}c@{}}$Q^{2}$ range\\$(\gev^{2})$\end{tabular}}} & \multicolumn{8}{c|}{NLO Predictions ($\pb$)}                                                                \\
\cline{3-10}
\multicolumn{2}{|c|}{}                  & \multicolumn{4}{c|}{FFN ABMP16.3}                     & \multicolumn{4}{c|}{FONLL-B NNPDF3.1}    \\
\cline{3-10}
\multicolumn{2}{|c|}{}                  & \multirow{2}{*}{$\sigma$} & \multicolumn{3}{c|}{uncertainties} & \multirow{2}{*}{$\sigma$} & \multicolumn{3}{c|}{uncertainties} \\
\cline{4-6}
\cline{8-10}
\multicolumn{2}{|c|}{}                  &                   & PDF & scale & mass &                   & PDF & scale & mass \\
\hline
\multicolumn{2}{|c|}{$e^+p$}                  & \multicolumn{4}{c|}{}                     & \multicolumn{4}{c|}{}                     \\
\hline
  200 & 1500    & $4.72$ & $\pm0.05$ & $^{+0.31}_{-0.23}$ & $\pm0.02$ &      $5.37$ &$\pm0.21$& $^{+0.68}_{-0.73}$ & $\pm0.00$     \\
1500 & 60000 &  $1.97$ & $\pm0.03$ & $^{+0.18}_{-0.13}$ & $\pm0.01$&       $2.66$ &$\pm0.23$& $^{+0.37}_{-0.26}$ & $\pm0.00$     \\
\hline
\multicolumn{2}{|c|}{$e^-p$}                  & \multicolumn{4}{c|}{}                     & \multicolumn{4}{c|}{}                     \\
\hline
  200 & 1500    & $4.50$ & $\pm0.05$ & $^{+0.31}_{-0.23}$ & $\pm0.02$ &      $4.98$ &$\pm0.22$& $^{+0.66}_{-0.71}$ & $\pm0.00$     \\
1500 & 60000 &  $1.73$ & $\pm0.03$ & $^{+0.18}_{-0.13}$ & $\pm0.01$  &      $2.16$ &$\pm0.22$& $^{+0.33}_{-0.21}$ & $\pm0.00$     \\
\hline
\end{tabular}
\caption{
  The NLO theory predictions from the FFN and FONLL-B schemes with their full uncertainties.
  The scale uncertainty was obtained by varying the renormalisation and factorisation scales
  simultaneously up and down by a factor two.
  The mass uncertainty was obtained by varying the charm mass, $m_c(m_c)$, within its uncertainties
  $m_c(m_c) = 1.28 \pm 0.03$~GeV.
}
\label{tab-prediction1}
\end{center}
\end{table}
\begin{table}[]
  \begin{center}
     \renewcommand{\arraystretch}{1.2}
\begin{tabular}{|c@{--}c|c|c|c|c|c|c|}
\hline
\multicolumn{2}{|c|}{\multirow{3}{*}{\begin{tabular}[c]{@{}c@{}}$Q^2$ range\\ {(}$\gev^2${)}\end{tabular}}} & \multicolumn{6}{c|}{NLO Predictions {(}$\pb${)}} \\
\cline{3-8}
\multicolumn{2}{|c|}{} & \multicolumn{5}{c|}{HERAPDF2.0}  & \multirow{2}{*}{\begin{tabular}[c]{@{}c@{}}ATLAS-\\ $epWZ$16\end{tabular}} \\
\cline{3-7}
\multicolumn{2}{|c|}{} & \begin{tabular}[c]{@{}c@{}}$f_s = 0.4$\\ (nominal)\end{tabular} & $f_s = 0.3$  & $f_s = 0.5$ & \begin{tabular}[c]{@{}c@{}}$f'_s = $\\ HERMES$^-$\end{tabular} & \begin{tabular}[c]{@{}c@{}}$f'_s = $\\ HERMES$^+$\end{tabular} &\\
\hline
\multicolumn{2}{|c|}{$e^+p$} & \multicolumn{6}{c|}{} \\
\hline
  200 & 1500    &$5.67$&$5.40$&$5.96$&$5.05$&$5.38$&$6.41$  \\
1500 & 60000 &$2.57$&$2.47$&$2.65$&$2.16$&$2.20$&$3.07$  \\
\hline
\multicolumn{2}{|c|}{$e^-p$}  & \multicolumn{6}{c|}{} \\
\hline
  200 & 1500    &$5.41$&$5.15$&$5.70$&$4.79$&$5.12$&$6.14$  \\
1500 & 60000 &$2.30$&$2.21$&$2.37$&$1.89$&$1.93$&$2.78$  \\
\hline
\end{tabular}
\caption{
  The NLO ZM-VFNS predictions with varied strange-quark fraction $f_s$.
  Additionally, two $x$-dependent strange quark fractions were used as suggested by the HERMES collaboration.
  The ZM-VFNS predictions were also evaluated with the ATLAS-$epWZ16$ PDF set with an unsuppressed strange-quark content.
}
\label{tab-prediction2}
\end{center}
\end{table}

\end{document}